%% file: eymh.tex
\documentclass{article}
\usepackage{amsmath,amsthm,amssymb,array,latexsym,cite,epsfig,finalT}

\numberwithin{equation}{section}

\newcommand{\sauthor}{T. Oliynyk}
\newcommand{\stitle}{Gravitating BPS monopole}

\input eymh.def

\begin{document}


\input eymh_tit.tex
\input intro.tex

\input feqns.tex

\input rsob.tex
\input esob.tex

\input sss.tex
\input mymh.tex
\input smooth.tex
\input solve.tex

\input exist.tex

\bigskip

\noindent \emph{Acknowledgements}.  
This work was partially supported
by the ARC grant A00105048 at the University of Canberra and
by the NSERC grants A8059 and 203614 at the University of Alberta.


\input eymh.bbl
\end{document}

%% file: eymh_tit.tex
\title{An existence proof for the gravitating BPS monopole
}
\author{
Todd A. Oliynyk \thanks{Present address: School of Mathematical Sciences, Monash University, VIC 3800 Australia}
\thanks{todd.oliynyk@sci.monash.edu.au} \\
Department of Mathematical and Statistical Sciences\\
University of Alberta\\ Edmonton AB T6G 2G1}

\date{}
\maketitle
\begin{abstract}
We prove the existence of the
gravitating BPS monopole in Einstein-Yang-Mills-Higgs (EYMH) theory.
Existence is established using a Newtonian perturbation argument which
shows  that a Yang-Mills-Higgs BPS monopole solution can be
be continued analytically in powers of $1/c^{2}$ to an EYMH solution.
\end{abstract}

%% file: intro.tex
\sect{intro}{Introduction}

In this paper we rigorously prove the existence of
the gravitating Bogomol'nyi-Prasad-Sommerfield (BPS) monopole 
which has been constructed numerically in \cite{BFM92}. 
We prove existence by using a Newtonian perturbation argument to show
that the flat space Yang-Mills-Higgs (YMH) BPS monopole  solution \cite{BPS} can be 
continued analytically 
to a Einstein-Yang-Mills-Higgs (EYMH) solution which we
refer to as the gravitating BPS monopole. 
The Newtonian perturbation argument in the form that is employed
in this paper was developed by Lottermoser in \cite{Lott92}
and subsequently used by Heilig to establish the existence
of slowly rotating stars \cite{Heil95}. 
For an elegant alternate presentation
of the Newtonian perturbation 
formalism using different but equivalent variables see
\cite{BS2000}. 

The results of Heilig and of this paper show that 
the Newtonian perturbation method is a powerful method for
obtaining existence theorems  in general relativity for static
or stationary matter models. In addition to establishing existence,
the method also provides an analytic deformation from
a Newtonian solution to its general relativistic counterpart.
The deformation parameter is $1/c^{2}$ where
$c$ is the speed of light. So a Taylor expansion  in  $1/c^2$ can
be considered as a converging post-Newtonian expansion. 
In this way, the Newtonian perturbation argument can be thought of
as the inverse of the Newtonian limit where Newtonian solutions
are obtained from general relativistic ones via the limit
$1/c^{2} \rightarrow 0$.  An attractive feature of the method is that 
it produces solutions  
to the Einstein field equations where the
matter fields are uniformly close to their corresponding
Newtonian solutions. This means that the properties of the Newtonian
solution pass directly to the corresponding relativistic solution.

In \cite{HD94} it is shown how to formulate the Newtonian limit of the
EYMH equations. The limiting equations have the important property that
the Newtonian potential and the YMH fields decouple.
Moreover, the static equations coincide with the static YMH equations
on Minkowski space. Since the BPS monopole is a static solution
to the YMH equations on Minkowski space, it can be interpreted
as a solution of the Newtonian YMH equations. Although
we use a different formalism from \cite{HD94}, the results are the
same. We find that in the limit as $1/c^{2}\rightarrow 0$, the
YMH variables decouple from the Newtonian potential
and also they satisfy the static YMH equations. This allows us to
use the BPS monopole solution as the starting point for the 
perturbation argument. Also, the fact that the Newtonian potential
decouples from the YMH variables in the limit $1/c^{2}\rightarrow 0$
helps to make the perturbation argument relatively simple.

The paper is organized as follows: in section \ref{feqns} we set up the field 
equations in a form
suitable to use the Newtonian perturbation argument while in section \ref{wsob}
we review the theory of weighted Sobolev spaces which will be
essential to our existence proof. The Banach spaces for our field
variables (i.e. the Higgs field, gauge potential, and metric density)
are set up in section \ref{sss} and then in section \ref{smooth} the
field equations are shown to be analytic on those spaces. Sections
\ref{solve}-\ref{exist} contain the Newtonian perturbation  argument.
In these sections it is shown that BPS monopole solution 
can be continued analytically to a solution of the full EYMH equations.

%


%% file: feqns.tex
\sect{feqns}{EYMH equations}

For indexing of tensors and related quantities
Greek indices, $\alpha,\beta,\gamma$ etc., will always
run from $0$ to $4$ while Roman indices, $i,j,k$ etc.,
will range from $1$ to $3$. Partial derivatives will be
denoted both by $\partial_{\alpha}u$ and
$u_{,\alpha}$ while covariant  derivatives
will be denoted by $\nabla_{\alpha}$.

Let $\underset{o}{g}$ denote the Minkowski metric on $\Rbb^{4}$.
Fix a global coordinate system $(x^{0},x^{1},
x^{2},x^{3})$ so that
\leqn{gflatup}{
\underset{o}{g}{}_{\alpha \beta} = \diag(-\lambda^{-1},1,1,1)
\quad \lambda := \frac{1}{c^{2}}
}
where $c$ is the speed of light. Define $\underset{o}{g}^{\alpha \beta}$
by $(\underset{o}{g}^{\alpha \beta}) := (\underset{o}{g}{}_{\alpha \beta})^{-1}$
which gives   
\leqn{gflatdwn}{
\underset{o}{g}{}^{\alpha \beta} = \diag(-\lambda,1,1,1) \, .
}
Define the Minkowski metric density
\leqn{godens}{
\go^{\alpha\beta} :=|\underset{o}{g}|^{\frac{1}{2}} 
\underset{o}{g}{}^{\alpha \beta} 
\quad \text{where} \quad |\underset{o}{g}| := 
|\det(\underset{o}{g}{}_{\alpha \beta})| \, .
}

Assume that $g_{\alpha \beta}$ is another Lorentzian metric defined
on $\Rbb^{4}$. Let $(g^{\alpha \beta}):= (g_{\alpha \beta})^{-1}$ and
introduce the density
\leqn{gdens}{
\g^{\alpha \beta} := |g|^{\frac{1}{2}} g^{\alpha\beta} 
\quad \text{where} \quad |g| := |\det(g_{\alpha \beta})| \, .
}
Following Lottermoser \cite{Lott92}, we form the tensor density
\leqn{udens}{
\U^{\alpha\beta} := \frac{1}{4\lambda^{\frac{3}{2}}}(\g^{\alpha \beta}
-\go^{\alpha \beta} ) \, 
}
which will be taken as our primary gravitational variable.
Observe that the metric
$g^{\alpha\beta}$ can be recovered from $\U^{\alpha\beta}$ by
\leqn{metfromdens1}{
g^{\alpha \beta} = \frac{1}{\sqrt{|g|}}\g^{\alpha\beta}
}
where
$\g^{\alpha \beta} = \go^{\alpha\beta} + 4\lambda^{\frac{3}{2}}\U^{\alpha\beta}$
and $|g| = |\det(\g^{\alpha\beta})|$.

The Einstein equations can be written in terms
of the density \eqref{udens} as \cite{Lott92},
\leqn{einst}{
4\pi G |\df| T^{\alpha \beta} = A^{\alpha\beta}
+ B^{\alpha \beta} + C^{\alpha\beta} + D^{\alpha \beta} \, ,
}
where
\lalign{einsta}{
\gob^{\alpha \beta} & := \sqrt{\lambda} \go^{\alpha \beta}   , \label{einsta1} \\
\gob{}_{\alpha \beta} & := \sqrt{\lambda} \go{}_{\alpha \beta} \quad \text{where}
\quad
(\go{}_{\alpha \beta}) := (\go^{\alpha \beta})^{-1}  ,\label{einsta2} \\
\gb^{\alpha \beta} & := \sqrt{\lambda} \g^{\alpha \beta} =
\gob^{\alpha \beta} + 4\lambda^{2} \U^{\alpha \beta}  , \label{einsta3}\\
\gb_{\alpha \beta} & := \sqrt{\lambda} \g_{\alpha \beta} \quad \text{where} \quad
(\g_{\alpha\beta}) := (\g^{\alpha \beta})^{-1} \, , \label{einsta4} \\
\df &:= \lambda \det(\g^{\alpha\beta})  ,\label{einsta5} \\
A^{\alpha \beta}& := 2\left(\Half \gb_{\mu\nu}
\gb_{\gamma\rho} - \gb_{\rho\mu}\gb_{\gamma\nu}\right)\left(
\gb^{\alpha\kappa}\gb^{\beta\sigma}- \Half \g^{\alpha \beta}\gb^{\kappa\sigma}\right)
\U^{\mu\nu}{}_{,\kappa}\U^{\gamma\rho}{}_{,\sigma}  ,\label{einsta6} \\
B^{\alpha\beta} & := 4\lambda \gb_{\kappa\sigma}\left(2\gb^{\gamma ( \alpha}
\U^{\beta )\sigma}{}_{,\rho}\U^{\kappa\rho}{}_{,\gamma}-\Half\gb^{\alpha\beta}
\U^{\kappa}{\rho}{}_{\gamma}\U^{\sigma\gamma}{}_{\rho}
-\gb^{\gamma\rho}\U^{\alpha\kappa}{}_{,\gamma}\U^{\beta\sigma}{}_{\rho}\right)
 , \label{einsta7} \\
C^{\alpha \beta} & := 4\lambda^{2}\left( \U^{\alpha\beta}{}_{,\kappa}
\U^{\kappa\rho}{}_{,\rho} - \U^{\alpha\kappa}{}_{,\rho} \U^{\beta\rho}{}_{,\kappa}
\right)  ,  \label{einsta8} \\
D^{\alpha \beta} & := \gb^{\mu\nu}\U^{\alpha \beta}{}_{,\mu\nu}+
\gb^{\alpha\beta}\U^{\mu\nu}{}_{,\mu\nu} -2\U^{\mu ( \alpha}{}_{,\mu\nu}
\gb^{\beta ) \nu}  ,  \label{einsta9} 
}
and $T^{\alpha \beta}$ is the stress-energy tensor. 
As discussed in \cite{Heil95}, any solution $(\lambda,\U^{\alpha\beta},T^{\alpha\beta})$
of \eqref{einst} for $\lambda >  0$ is a solution of 
Einstein's equations displayed in units where $c=1/\sqrt{\lambda}$.
Following \cite{Heil95},
we  choose harmonic coordinates
\eqn{harmonic}{
\nabla_{\alpha}\nabla^{\alpha} x^{\beta} = 0 \, , \quad \text{ or equivalently}
\quad \U^{\alpha \beta}{}_{,\beta} = 0 \, ,
}
which allows us to write the full Einstein field equations as
\lgath{reduced}{
\U^{\alpha \beta}{}_{,\beta} = 0 \, ,\label{reduced2} \\
4\pi G |\df| T^{\alpha \beta} = E^{\alpha \beta} \, ,\label{reduced3}
}
where
\lgath{reduceda}{
E^{\alpha \beta}  := \gob^{\mu\nu}\U^{\alpha\beta}{}_{,\mu\nu} + 
4\lambda^{2}\left(\U^{\mu\nu}\U^{\alpha\beta}{}_{,\mu\nu} + 
\U^{\alpha\beta}\U^{\mu\nu}{}_{,\mu\nu} - 2\U^{\mu (\alpha}{}_{,\mu\nu}
\U^{\beta ) \nu}\right) \notag \\
 + A^{\alpha\beta} + B^{\alpha\beta} + C^{\alpha\beta} \label{reduceda1} \, .
}
The equations \eqref{reduced3} will be called the \emph{reduced field equations}.

It is important to recognize that alone the reduced field equations \eqref{reduced3}
are not equivalent to the Einstein field equations \eqref{einst}.
However, it is shown in \cite{Heil95} \S 6  that if $\nabla_{\beta}T^{\alpha \beta}=0$
 and \eqref{reduced3}
can be solved and the stress-energy tensor $T^{\alpha \beta}$ satisfies
certain conditions then the harmonic condition \eqref{reduced2} will be
automatically satisfied. In this case, a solution to \eqref{reduced3} will
actually be a solution to the full Einstein equation \eqref{einst}.

We will let
$A = A_{\alpha} dx^{\alpha}$
denote the $SU(2)$-gauge potential and
$\Phi$ the Higgs field. 
The SU(2) Yang-Mills-Higgs equations are
\lgath{YMH}{
g^{\nu\alpha}D^{A}_{\nu} F^{A}_{\alpha \beta} = [\Phi,D^{A}_{\beta}\Phi] \, ,\label{YMH1}\\
g^{\nu\alpha}D^{A}_{\nu} D^{A}_{\alpha}\Phi = 0 \, , \label{YMH2}
}
where 
\leqn{gder}{
D^{A}_{\alpha}(\cdot) := \nabla_{\alpha}(\cdot)+ [A_{\alpha},\cdot]
}
is the gauge covariant derivative on gauge-scalars and
\leqn{gfield}{
F^{A}_{\alpha \beta} := \partial_{\alpha}A_{\beta} - \partial_{\beta}A_{\alpha} +
[A_{\alpha}, A_{\beta}]
}
is the gauge field. For later use we define 
\leqn{fgder}{
\Do{}^{A}_{\alpha}(\cdot) := \partial_{\alpha}(\cdot)+ [A_{\alpha},\cdot]
}
which is the gauge covariant derivative on Minkowski space.

Multiplying \eqref{YMH1} and \eqref{YMH2} by $\sqrt{\lambda|g|}$ 
we find that
\lgath{YMHa}{
\gb^{\alpha\nu}\left(F_{\alpha\beta,\nu} - \Gamma^{\mu}_{\alpha\nu}
F_{\mu\beta} - \Gamma^{\mu}_{\beta\nu} F_{\alpha\mu} +
[A_{\nu},F_{\alpha\beta}]\right) - \sqrt{\df}[\Phi,D_{\beta}\Phi] =0
\, , \label{YMHa1} \\
\gb^{\alpha\beta}\left(\partial_{\alpha}D_{\beta}\Phi  - 
\Gamma^{\mu}_{\alpha\beta}D^{A}_{\mu}\Phi
+[A_{\alpha},D_{\beta}\Phi]\right) = 0 \, , \label{YMHa2} 
} 
where the Christoffel $\Gamma^{\alpha}_{\beta\gamma}$ symbols are given by
\leqn{Christ}{
\Gamma^{\alpha}_{\beta\gamma} = 
\gb^{\alpha\mu}(2\gb_{\beta\sigma}\gb_{\gamma\tau} - \gb_{\beta\gamma}
\gb_{\sigma\tau})\U^{\sigma\tau}{}_{,\mu} + 2\lambda (
\gb_{\sigma\tau}\delta^{\alpha}_{(\beta}\U^{\sigma\tau}{}_{,\gamma)}-
2\gb_{\sigma(\beta}\U^{\alpha\sigma}{}_{,\gamma)} ) \, .
}
We note that since $\Phi$ is a $\g$-valued scalar,
\leqn{DPhi}{
D^{A}_{\alpha}\Phi = \partial_{\alpha}\Phi
+[A_{\alpha},\Phi]
}
does not involve the metric.

The stress-energy tensor can be written as
\lalign{streng}{
T^{\alpha \beta} & = \left( g^{\alpha\mu}g^{\beta\nu}
\ip{D^{A}_{\mu}\Phi}{D^{A}_{\nu}\Phi} - \Half g^{\alpha\beta}g^{\mu\nu}
\ip{D^{A}_{\mu}\Phi}{D^{A}_{\nu}\Phi}
\right) + \notag \\
& \left( g^{\alpha\mu}g^{\beta\nu}g^{\sigma\tau}
\ip{F^{A}_{\mu\sigma}}{F^{A}_{\nu\tau}}-\Quarter g^{\mu\nu}g^{\sigma\tau}g^{\alpha\beta}
\ip{F^{A}_{\mu\sigma}}{F^{A}_{\nu\tau}} \right) \label{streng1} \, .
}
where $\ip{\cdot}{\cdot}$ is an $\Ad$-invariant positive definite
inner-product on $\sU{2}$.
Using the YMH equations \eqref{YMH1}-\eqref{YMH2}
, it is straightforward
to verify that any YMH solution satisfies   
\leqn{streng1a}{
\nabla_{\beta} T^{\alpha \beta} = 0 
}
automatically irrespective of the metric. Consequently,
it will be enough to solve the reduced field equations \eqref{reduced3}
and the YMH equations \eqref{YMH1}-\eqref{YMH2}
to obtain a solution to the full EYMH field equations. 

Let
\leqn{streng2}{
\Tc^{\alpha\beta} := 4\pi G|\df| T^{\alpha\beta}
}
so that
\lalign{streng3}{
\Tc^{\alpha \beta}& = 4\pi G  
\left( \gb^{\alpha\mu}\gb^{\beta\nu}
\ip{D^{A}_{\mu}\Phi}{D^{A}_{\nu}\Phi} - \Half \gb^{\alpha\beta}\gb^{\mu\nu}
\ip{D^{A}_{\mu}\Phi}{D^{A}_{\nu}\Phi}
\right)
+ \notag \\
& \frac{4\pi G}{\sqrt{|\df|}}\left( \gb^{\alpha\mu}\gb^{\beta\nu}
\gb^{\sigma\tau}
\ip{F^{A}_{\mu\sigma}}{F^{A}_{\nu\tau}}-\Quarter \gb^{\mu\nu}\gb^{\sigma\tau}\gb^{\alpha\beta}
\ip{F^{A}_{\mu\sigma}}{F^{A}_{\nu\tau}} \right) \label{streng4} \, .
}

%% file: rsob.tex
\sect{wsob}{Weighted Sobolev Spaces}

In this section we introduce two different types of weighted Sobolev
spaces and prove a number of results that will be essential to our
existence proof. The following subsets of $\Rbb^{n}$  will be needed: $B_{R}(x)$ the open ball of radius $R$
centered at $x\in \Rbb^{n}$, $Q_{R}(x)$ the open n-cube centered at $x$ with vertices defined
by the boundary of $B_{R}(x)$, and the exterior domain $E_{R}(x) :=\Rbb^{n}\setminus B_{R}(x)$.
We will also repeatedly use the  cutting function $\chi_{R}\in C^{\infty}_{0}(\Rbb^{n})$ which
is defined as follows: let $\chi\in C^{\infty}{[0,\infty)}$ be any function such that
\leqn{chi1}{
\chi|_{[0,1)} = 1,\quad \text{supp}\chi \subset [0,2), \quad \text{and} \quad
0\leq \chi \leq 1\, .
}
Then for $R > 0$, $\chi_{R}$ is given by
\leqn{chi2}{
\chi_{R}(x) := \chi(\enorm{x}/R) \, .
}
 
\subsect{rsob}{Radially weighted Sobolev Spaces}

Let $V$ denote a finite dimensional vector space with
norm $\enorm{\cdot}$. 
\begin{Def}  \label{rsobdef1} \mnote{[rsobdef1]}
The \emph{\rad weighted Lesbegue space} $\text{L}^{p}_{\delta}(\Rbb^{n},V)$,
$1\leq p \leq \infty$, with weight $\delta \in \Rbb$ is the set
of all measurable maps from $\Rbb^{n}$ to $V$ in 
$\text{L}^{p}_{\text{loc}}(\Rbb^{n},V)$ such that the norm
\eqn{rLebnorm}{
\norm{u}_{p,\delta} = \left\{ \begin{array}{ll}
{\displaystyle \left(\int_{\Rbb^{n}}|u|^{p}\sigma^{-\delta p -n} d^{n}x\right)^
{\frac{1}{p}}} &
\text{if $p <\infty$}\, \\
&  \\
\text{\emph{ess sup}}_{\Rbb^{n}}(\sigma^{-\delta}|u|) & \text{if $p=\infty$} \, ,
\end{array} \right.
}
is finite. Here $\sigma(x) := \sqrt{|x|^{2} + 1}$. If $V=\Rbb$
then we write $\text{L}^{p}_{\delta}(\Rbb^{n})$ instead
of $\text{L}^{p}_{\delta}(\Rbb^{n},V)$.
\end{Def}

\begin{Def}  \label{rsobdef2} \mnote{[rsobdef2]}
The \emph{\rad weighted Sobolev space} $\W^{k,p}_{\delta}(\Rbb^{n},V)$,
$1\leq p \leq \infty$, $k\in\Nbb_{0}$, with weight $\delta \in \Rbb$ is the set
\eqn{rSobelev}{
\W^{k,p}_{\delta}(\Rbb^{n},V):= \{\, u \in 
\text{L}^{p}_{\delta}(\Rbb^{n},V) \,|\,
\partial^{I}u\in \text{L}^{p}_{\delta-|I|}(\Rbb^{n},V) \;\text{for all
$I : |I|\leq k$}\,\}
}
with norm
\eqn{rSobnorm}{
\norm{u}_{k,p,\delta} := \sum_{|I|\leq k} \norm{\partial^{I}u}_{p,\delta-|I|}\, ,
}
where $I=(I_{1},I_{2},\ldots,I_{n})$ is a multi-index
and $\partial^{I} :=\partial_{1}^{I_{1}}\partial_{2}^{I_{2}}\cdots\partial_{n}^{I_{n}}$.
If $V=\Rbb$ then we will write $\W^{k,p}_{\delta}(\Rbb^{n})$ instead
of $\W^{k,p}_{\delta}(\Rbb^{n},V)$.
\end{Def}
We note that the set $\Co(\Rbb^{n},V)$ of smooth maps
from $\Rbb^{n}$ to $V$ with compact support is dense in $\W^{k,p}_{\delta}(\Rbb^{n},V)$.
As above, if $V=\Rbb$ then we write $\Co(\Rbb^{n})$ instead of $\Co(\Rbb^{n},V)$

Two easy consequences of these definitions are that differentiation
\leqn{rdiff}{
\partial_{j} : \W^{k,p}_{\delta}(\Rbb^{n},V) \longrightarrow 
\W^{k-1,p}_{\delta-1}(\Rbb^{n},V) \; : \; u \longmapsto \partial_{j}u
}
is a continuous map and that 
\leqn{rincl}{
\W^{k,p}_{\delta_{2}}(\Rbb^{n},V) \subset \W^{k,p}_{\delta_{1}}(\Rbb^{n},V)
 \quad \text{for $\delta_{2}\leq \delta_{1}$. }
}

As with the Sobolev spaces, we can define weighted versions 
of the  
\eqn{ckb}{
C^{k}_{B}(\Rbb^{n},V):=C^k(\Rbb^n,V)\cap W^{k,\infty}(\Rbb^n,V)} 
and $C^{k,\alpha}(\Rbb^{n},V)$ spaces.
For a map $u\in C^{0}(\Rbb^{n},V)$ and $\delta\in \Rbb$, $\alpha > 0$ , let
\eqn{rsobdef3.1}{
\norm{u}_{C^{0}_{\delta}} := \sup_{x\in \Rbb^{n}}|\sigma(x)^{-\delta}u(x)| 
}
and
\eqn{rsobdef3.2}{
\norm{u}_{C^{0,\alpha}_{\delta}}  := \norm{u}_{C^{0}_{\delta}}
+ \sup_{x\in \Rbb^{n}}\Bigl(\sigma^{-\delta+\alpha}(x) \sup_{4|x-y|\leq \sigma(x)}\frac{|u(x)-u(y)|}{|x-y|^{\alpha}}
\Bigr)\, .
}
Using these two norms  we define the norms $\norm{\cdot}_{C^{k}_{\delta}}$ and
$\norm{\cdot}_{C^{k,\alpha}_{\delta}}$ in the usual way:
\eqn{rsobdef3.3}{
\norm{u}_{C^{k}_{\delta}} := \sum_{|I|\leq k} \norm{\partial^{I}u}_{C^{0}_{\delta-|I|}} \,  
} 
and
\eqn{rsobdef3.4}{
\norm{u}_{C^{k,\alpha}_{\delta}} := \sum_{|I|\leq k} \norm{\partial^{I}u}_{C^{0,\alpha}_{\delta-|I|}} \, .
}
So then
\eqn{rsobdef3.5}{
 C^{k}_{\delta}(\Rbb^{n},V) := \bigl\{\, u \in C^{k}(\Rbb^n,V) \, | \, \norm{u}_{C^{k}_{\delta}} < \infty \: \bigr\}
}
and
\eqn{rsobdef3.6}{
 C^{k,\alpha}_{\delta}(\Rbb^{n},V) := 
\bigl\{\, u \in C^{k}(\Rbb^n,V) \, | \, \norm{u}_{C^{k,\alpha}_{\delta}} < \infty \: \bigr\}\, .
}

Our main references for the \rad weighted Sobolev spaces will be
\cite{Bart86} and \cite{CHCH81}. Contained in these articles
are a number useful theorems including weighted versions of
the Sobolev embedding theorems, the Rellich-Kondrachov theorem,
and interior estimates for elliptic operators.
Also contained in these papers in an analysis of the Laplace
operator and its mapping properties between the \rad  weighted
spaces. We will frequently 
require results from these papers and will refer the reader to the 
appropriate theorems. A result we would like to mention is the following 
improvement of lemma 2.5  of  \cite{CHCH81}.
\begin{lem}  \label{multiply} \mnote{[multiply]}
If there exists a multiplication 
$V_{1}\times V_{2} \rightarrow V_{3}$ $(u,v)\mapsto u\cdot v$ then 
for $1\leq p < \infty$ the corresponding multiplication
\eqn{multiply1}{
\W^{k_{1},p}_{\delta_{1}}(\Rbb^{n},V_{1})\times \W^{k_{2},p}_{\delta_{2}}(\Rbb^{n},V_{2}) 
\rightarrow \W^{k_{3},p}_{\delta_{3}}(\Rbb^{n},V_{3})\; : \; (u,v) 
\mapsto u\cdot v
}
is bilinear and continuous if
$k_{1},k_{2} \geq k_{3}$, $k_{3} < k_{1} + k_{2} - n/p$, and
$\delta_{1} +\delta_{2} \leq \delta_{3}$ .
\end{lem}
\begin{proof} This can be proved using the weighted Sobolev and
H\"{o}lder inequalities from theorem 1.2 of \cite{Bart86} in exactly
the same fashion as for the regular unweighted Sobolev 
spaces. Note that
theorem 1.2 of \cite{Bart86} is missing the weighted
version of the Sobolev inequality for $kp=n$. The same arguments
in theorem 1.2 can be used to establish this case which reads:
if $u \in \W^{k,p}_{\delta}$ and $n=kp$, then
$\norm{u}_{q,\delta} \leq C\norm{u}_{p,k,\delta}$ for
$p\leq q < \infty$. 
\end{proof}
We also will need the following variation of proposition 1.6 of \cite{Bart86}.
\begin{prop} \label{eestimate} \mnote{[eestimate]}
Suppose $1 < p  < \infty$ and $\delta \in \Rbb$ and
$f(x)$ is a continuous function that satisfies
$f(x)=\Ord(|x|^{-2})$ as $|x|\rightarrow \infty$.
Then there exists a constant $C$ such that if $u\in L^{0,p}_{\delta}$
and $\Delta u + fu \in L^{0,p}_{\delta-2}$ then
$u \in W^{2,p}_{\delta}$ and
\eqn{eestimate.1}{
\norm{u}_{2,p,\delta} \leq C\bigl(\norm{\Delta u+fu}_{0,p,\delta-2}
+ \norm{u}_{0,p,\delta}\bigr) \, .
}
\end{prop}
\begin{proof}
This proof follows from the local elliptic estimates and scaling in exactly
the same fashion as the proof of proposition 1.6 in \cite{Bart86}.
\end{proof}

%% file: esob.tex
\subsect{esob}{Exponentially weighted Sobolev Spaces}

\begin{Def}  \label{esobdef1} \mnote{[esobdef1]}
The \emph{exponentially weighted Lesbegue space} $\Lc^{p}_{\mu}(\Rbb^{n},V)$,
$1\leq p \leq \infty$, with weight $\mu \in \Rbb$ is the set
of all measurable maps from $\Rbb^{n}$ to $V$ in 
$\Lc^{p}_{\text{loc}}(\Rbb^{n},V)$ such that the norm
\eqn{ewnorm}{
\nnorm{u}_{p,\mu} = \left\{ \begin{array}{ll}
{\displaystyle \left(\int_{\Rbb^{n}}|u(x)|^{p}e^{-\mu p\enorm{x}} d^{n}x\right)^
{\frac{1}{p}}} &
\text{if $p <\infty$}\, \\
&  \\
\text{\emph{ess sup}}_{\Rbb^{n}}(e^{-\mu|x|}|u(x)|) & \text{if $p=\infty$} \, ,
\end{array} \right.
}
is finite. If $V=\Rbb$
then we write $\Lc^{p}_{\mu}(\Rbb^{n})$ instead
of $\Lc^{p}_{\mu}(\Rbb^{n},V)$.
\end{Def}

\begin{Def}  \label{esobdef2} \mnote{[esobdef2]}
The \emph{exponentially weighted Sobolev space} $\Wc^{k,p}_{\mu}(\Rbb^{n},V)$,
$1\leq p \leq \infty$, $k\in\Nbb_{0}$, with weight $\mu \in \Rbb$ is the set
\eqn{eSobelev}{
\Wc^{k,p}_{\mu}(\Rbb^{n},V):= \{\, u \in 
\Lc^{p}_{\mu}(\Rbb^{n},V) \,|\,
\partial^{I}u\in \Lc^{p}_{\mu}(\Rbb^{n},V) \;\text{for all
$I : |I|\leq k$}\,\}
}
with norm
\eqn{eSobnorm}{
\nnorm{u}_{k,p,\mu} := \sum_{|I|\leq k} \nnorm{\partial^{I}u}_{p,\mu}\, .
}
If $V=\Rbb$ then we will write $\Wc^{k,p}_{\delta}(\Rbb^{n})$ instead
of $\Wc^{k,p}_{\mu}(\Rbb^{n},V)$.
\end{Def}
We note that the $\Co(\Rbb^{n},V)$ 
is dense in $\Wc^{k,p}_{\mu}(\Rbb^{n},V)$. 
A straightforward consequence of the above definitions is that differentiation
\leqn{ediff}{
\partial_{j} : \Wc^{k,p}_{\mu}(\Rbb^{n},V) \longrightarrow \Wc^{k-1,p}_{\mu}(\Rbb^{n},V) \; : \; u \longmapsto \partial_{j}u
}
is a continuous map. Also note that $\Wc^{k,p}_{0}(\Rbb^{n},V)
 = \W^{k,p}(\Rbb^{n},V)$ while it follows from \cite{Bart86} theorem 1.2 (i)
that $\W^{k,p}_{\delta}(\Rbb^{n},V) \subset \W^{k,p}(\Rbb^{n},V)$ for $\delta \leq -n/p$.
Consequently we have the inclusion
\leqn{reinc}{
\W^{k,p}_{\delta}(\Rbb^{n},V) \subset \Wc^{k,p}_{0}(\Rbb^{n},V) \quad \text{for $\delta \leq -n/p$.}
}
It also follows directly from H\"{o}lders inequality and 
the definitions of the \rad and exponentially weighted spaces that
\leqn{mudel}{
\Wc^{k,p}_{\mu}(\Rbb^{n},V) \subset \W^{k,p}_{\delta}(\Rbb^{n},V) \quad \text{for all $\delta \in \Rbb$ provided $\mu < 0$.}
}

As with the \rad weighted case, we can also define the corresponding exponential weighted
$C^{k}_{B}(\Rbb^{n},V)$ and $C^{k,\alpha}(\Rbb^{n},V)$ spaces.
For a map $u\in C^{0}(\Rbb^{3},V)$ and $\delta\in \Rbb$, $\alpha > 0$ , let
\eqn{esobdef3.1}{
\norm{u}_{\Cc^{0}_{\mu}} := \sup_{x\in \Rbb^{n}}|e^{-\mu|x|}u(x)|
}
and
\eqn{esobdef3.2}{
\norm{u}_{\Cc^{0,\alpha}_{\mu}}  := \norm{u}_{\Cc^{0}_{\mu}}
+ \sup_{x\in \Rbb^{n}}\Bigl(e^{-\mu|x|}\sup_{|x-y|\leq 1}\frac{|u(x)-u(y)|}{|x-y|^{\alpha}}
\Bigr)\, .
}
Using these two norms  we define the norms $\norm{\cdot}_{\Cc^{k}_{\mu}}$ and
$\norm{\cdot}_{\Cc^{k,\alpha}_{\mu}}$ by
\eqn{esobdef3.3}{
\norm{u}_{\Cc^{k}_{\mu}} := \sum_{|I|\leq k} \norm{\partial^{I}u}_{\Cc^{0}_{\mu}} \,
}
and
\eqn{esobdef3.4}{
\norm{u}_{\Cc^{k,\alpha}_{\mu}} := \sum_{|I|\leq k} \norm{\partial^{I}u}_{\Cc^{0,\alpha}_{\mu}} \, .
}
So then
\eqn{esobdef3.5}{
 \Cc^{k}_{\mu}(\Rbb^{n},V) := \bigl\{\, u \in C^{k}(\Rbb,V) \, | \, \norm{u}_{\Cc^{k}_{\mu}} < \infty \: \bigr\}
}
and
\eqn{esobdef3.6}{
 \Cc^{k,\alpha}_{\mu}(\Rbb^{n},V) :=
\bigl\{\, u \in C^{k}(\Rbb,V) \, | \, \norm{u}_{\Cc^{k,\alpha}_{\mu}} < \infty \: \bigr\}\, .
}

To prove weighted versions of the Sobolev inequalities from local inequalities, a covering
argument is needed. Let  $\{x_{a}\}_{a\in\Zbb^{n}}$ be a sequence of points 
such that
\leqn{cdecomp}{
\Rbb^{n} = \bigcup_{a\in\Zbb^{n}} \overline{Q_{R}(x_{a})}
}
and  $Q_{R}(x_{a})\cap Q_{R}(x_{a'})=\emptyset$ for $a\neq a'$. Then
there exists a number $N$ independent of $a$  such that the set  
\leqn{cdcomp1}{
\{\, a'\in \Zbb^{n}\,|\,B_{2R}(x_{a})\cap Q_{R}(x_{a'}) \neq \emptyset\,\} 
}
has at most $N$ elements. The key property we need is that for any $\sigma \in \Rbb$
there exists a constant $C=C(\sigma,R)$ independent of $x\in \Rbb^{n}$ such that
\leqn{cdecomp2}{ 
C^{-1}e^{\sigma|x|}\leq e^{\sigma|y|} \leq C e^{\sigma|x|} \qquad \forall y \in B_{R}(x)\,.
}
From this inequality it follows that there
exists a constant $C$ independent of $x$ such that
\leqn{cdecomp3}{
C^{-1}e^{-\mu|x|}\norm{u_{x}}_{k,p;B_{R}(0)}  \leq \nnorm{u}_{k,p,\mu;B_{R}(x)}
\leq Ce^{-\mu|x|} \norm{u_{x}}_{k,p;B_{R}(0)}
}
where 
\leqn{cdecomp4}{
u_{x}(y) := u(x+y).
}
Note that the constant only depends on $\mu, p, k$ and
$R$. Equations \eqref{cdecomp}-\eqref{cdecomp4} will allow us to turn local estimates
into global ones. The next theorem generalizes the H\"{o}lder and Sobolev inequalities
to the exponentially weighted spaces and the proof closely follows that
of theorem 1.2 of \cite{Bart86}. 

\begin{thm} \label{esobthm1} \mnote{[esobthm1]}
\begin{itemize}
\item[(i)] If $1\leq p \leq q\leq \infty$, $\mu_{2}<\mu_{1}$
and $u \in \Lc^{q}_{\mu_{2}}$, then
\eqn{esobthm1.1}{
\nnorm{u}_{p,\mu_{1}} \leq C\,\nnorm{u}_{q,\mu_{2}}
}
and hence $\Lc^{q}_{\mu_{2}} \subset \Lc^{p}_{\mu{1}}$.
\item[(ii)] If $1\leq p,q,r\leq \infty$, $\frac{1}{r} = \frac{1}{q}+\frac{1}{p}$,
$u\in \Lc^{q}_{\mu_{1}}$, $v\in \Lc^{p}_{\mu_{2}}$, and $\mu_{3} = \mu_{1}+\mu_{2}$
then
\eqn{esobthm1.2}{
\nnorm{uv}_{r,\mu_{3}} \leq \nnorm{u}_{q,\mu{1}}\nnorm{v}_{p,\mu_{2}} \, .
}
\item[(iii)] For any $\epsilon > 0$, there is a $C(\epsilon)$ such that for
all $u\in \Wc^{2,p}_{\mu}$, $1\leq p \leq \infty$,
\eqn{esobthm1.2a}{
\nnorm{u}_{1,p,\mu} \leq \epsilon \nnorm{u}_{2,p,\mu} + C(\epsilon)
\nnorm{u}_{0,p,\mu}.
}

\item[(iv)] If $u\in \Wc^{k,p}_{\mu}$ and $n-kp > 0$ then
\eqn{esobthm1.3}{
\nnorm{u}_{q,\mu} \leq C\nnorm{u}_{k,p,\mu}
}
for $p\leq q\leq np/(n-kp)$.
\item[(v)] If $u\in \Wc^{k,p}_{\mu}$ and $n-kp = 0 $ then
\eqn{esobthm1.4}{
\nnorm{u}_{q,\mu} \leq C\nnorm{u}_{k,p,\mu}
}
for $p\leq q < \infty$.
\item[(vi)]  If $u\in \Wc^{k,p}_{\mu}$ and $n- kp < 0$ then
$u \in \Cc^{0}_{\mu}$ and
\eqn{esobthm1.4a}{
\nnorm{u}_{\Cc^{0}_{\mu}} \leq C\nnorm{u}_{k,p,\mu} \, .
}
Moreover $|u(x)|=\ord(e^{\mu |x|})$ as $|x|\rightarrow \infty$.
\end{itemize}
\item[(vii)]
If $u\in \Wc^{k,p}_{\mu}$, $0 < \alpha \leq k-n/p \leq 1$, then
$u \in \Cc^{0,\alpha}_{\mu}$ and 
\eqn{esobthm1.4b}{
\nnorm{u}_{\Cc^{0,\alpha}_{\mu}} \leq C\nnorm{u}_{k,p,\mu} \, .
}
\end{thm}
\begin{proof}
Parts (i) and (ii) follow from the definition and H\"{o}lder's inequality. 
The proofs of (iii)-(vii) follow from the interpolation and Sobolev inequalities on $B_{1}(0)$
together with equations  \eqref{cdecomp}-\eqref{cdecomp4}. We will only prove (iv) and
leave the remainder to the reader. So assume that $n-pk>0$, $p \leq q \leq np/(n-kp)$,
and $u\in \Wc^{k,p}_{\mu}$. Then clearly $u_{x} \in \W^{k,p}(B_{2}(0))$ and hence applying
the standard Sobolev inequality yields
\eqn{esobthm1.5}{
\norm{u_{x}}_{q;B_{2}(0)} \leq C\norm{u_{x}}_{p,k;B_{2}(0)}= C\norm{u}_{p,k;B_{2}(x)}\, .
}
The constant $C$ above only depends on $p$, $k$, and the ball $B_{2}(0)$. Using
\eqref{cdecomp3} we get
\leqn{esobthm1.6}{
\nnorm{u}_{q,\mu;B_{2}(x)} \leq C \nnorm{u}_{p,k,\mu;B_{2}(x)} 
}
for a constant $C$ independent of $u$ and $x$. So 
\lalign{esobthm1.7}{
\nnorm{u}_{q,\mu}
& \leq \left(\sum_{a\in \Zbb^{n}}\nnorm{u}^{q}_{q,\mu;B_{2}(x_{a})} \right)^{1/q} 
\notag\\
& \leq C\left(\sum_{a\in \Zbb^{n}}\nnorm{u}^{q}_{p,k,\mu;B_{2}(x_{a})} \right)^{1/q}
& \text{by \eqref{esobthm1.6}} \notag \\
& \leq C\left(\sum_{a\in \Zbb^{n}}\nnorm{u}^{p}_{p,k,\mu;B_{2}(x_{a})} \right)^{1/p}
\label{esobthm1.7.1}
}
where in deriving the last inequality we have used $(\sum_{j} b_{j}^{s})^{1/s} 
\leq (\sum_{j} b_{j}^{t})^{1/t}$ for $b_{j}\geq 0$ and $t\leq s$. Using the
finite intersection property \eqref{cdcomp1}, there exists a constant
$K$ independent of $u$ such that
\leqn{esobthm1.8}{
\sum_{a\in \Zbb^{n}}\nnorm{u}^{p}_{p,k,\mu;B_{2}(x_{a})} \leq K \nnorm{u}^{p}_{p,k,\mu} \,.
}
To see this it is enough to show it for the norm $\nnorm{\cdot}_{p,\mu}$. From
the finite interesction property we know that there exists a set of
points $\{x_{a_1}=x_a,x_{a_2},\ldots,x_{a_N}\}$ such that
\eqn{cov1}{
B_2(x_{a}) \subset E \cup \bigcup_{j=1}^{N} Q_1(x_{a_j}) 
}
where $E$ is a set of measure zero. So
\alin{cov2}{
\nnorm{u}_{p,\mu;B_2(x_a)}^p &= \int_{B_2(x_a)} 
|u(x)|^p e^{-\mu p |x|}d^n x 
\leq \sum_{j=1}^{N}  \int_{Q_1(x_{a_j})} 
|u(x)|^p e^{-\mu p |x|}d^n x \\
}
and hence
\alin{cov3}{ 
\sum_{a\in \Zbb^{n}}&\nnorm{u}^{p}_{p,\mu;B_{2}(x_{a})}
\leq \sum_{a\in \Zbb^{n}}  \sum_{j=1}^{N}  \int_{Q_1(x_{a_j})}
|u(x)|^p e^{-\mu p |x|}d^n x \\
& \leq N \sum_{a\in \Zbb^{n}} \int_{Q_1(x_{a})}
|u(x)|^p e^{-\mu p |x|}d^n x 
 = N\int_{\Rbb^n}|u(x)|^p e^{-\mu p |x|}d^n x = N\nnorm{u}_{p,\mu} \, . 
}
Note in getting the second to last equality we use the fact that the
set 
\eqn{cov4}{
\Rbb^n\setminus\bigl(\bigcup_{a\in \Zbb^n}Q_{1}(x_a)\bigr)
}
has measure zero. Equation \eqref{esobthm1.8} now follows immediately.
Combining \eqref{esobthm1.7.1} and \eqref{esobthm1.8} proves (iii).
\end{proof}

The first of the following two lemmas is the exponentially weighted version of the
Rellich-Kondrachov theorem  and both lemmas can be proved  by adapting
the proof of lemma 2.1 in \cite{CHCH81}. We only prove the second
and leave the first to the reader.

\begin{lem} \label{esoblem2} \mnote{[esoblem2]}
For $k_{1} > k_{2}$, $\mu_{1} < \mu_{2}$ and $1\leq p < \infty$ the
inclusion $\Wc^{k_{1}}_{\mu_{1}} \subset \Wc^{k_{2}}_{\mu_{2}}$
is compact.
\end{lem}
\begin{lem} \label{esoblem3} \mnote{[esoblem3]}
Suppose $v\in W^{\infty,k_{1}}(\Rbb^{n})$ and the function
\eqn{esoblem3.1}{
\xi(R) := \underset{0\leq |I|\leq k_{1}}{\max}\underset{x\in E_{R}(0)}{\sup}
|\partial^{I} v(x)| 
}
satisfies $\lim_{R\rightarrow \infty} \xi(R) = 0$. Then for $k_{1} > k_{2}$ and
$1\leq p < \infty$ the map
\eqn{esoblem3.2}{
\Wc^{k_{1}}_{\mu} \longrightarrow \Wc^{k_{2},p}_{\mu}\: : \: u \longmapsto vu
}
is compact. 
\end{lem} 
\begin{proof}
Let $\{u_{n}\} \in \Wc^{k_{1},p}_{\mu_{1}}$ be a sequence such that 
$\nnorm{u_{n}}_{k_{1},p,\mu}\leq 1$. Then there exists a subsequence still denoted
$\{u_{n}\}$ such that $u_{n}\rightarrow u$ weakly in $\Wc^{k_{1},p}_{\mu}$ for
some $u\in \Wc^{k_{1},p}_{\mu}$ with $\nnorm{u}_{k_{1},p,\mu}\leq 1$.
From theorem \ref{esobthm1} (ii) we have that  $\nnorm{vu}_{k_{1},p,\mu} \leq C\nnorm{u}_{k_{1},p,\mu}$
for some $C$ that depends only on $\xi(R)$. Therefore the map
\leqn{esoblem3.3}{
L_{v} : \Wc^{k_{1},p}_{\mu}\longrightarrow \Wc^{k_{1},p}_{\mu} ;\; u \longmapsto vu
}
is continuous and hence weakly continuous. So $vu_{n} \rightarrow vu$ weakly
in $\Wc^{k_{1},p}_{\mu}$. By \eqref{cdecomp2}
there exist a constant $C_{R}$ depending only
on $\mu$, $p$ and $\norm{\chi_{R}}_{C^{k_{1}}(B_{R}(0))}$ such that
\leqn{esoblem3.4}{
\norm{\chi_{R}vu_{n}}_{k_{1},p;B_{2R}(0)} \leq C_{R}\nnorm{vu_{n}}_{k_{1},p,\mu} \, .
}
But then 
\eqn{esoblem3.5}{
\norm{\chi_{R}vu_{n}}_{k_{1},p;B_{2R}(0)} \leq C_{R}\norm{L_{v}}_{\op}
}
by \eqref{esoblem3.3}, \eqref{esoblem3.4}, and $\nnorm{u_{n}}_{k_{1},p,\mu} \leq 1$,
where $\norm{L_{v}}_{\op}$ denotes the operator norm of $L_{v}$
The compactness of the embedding
$\W^{k_{1},p}(B_{2R}(0))\rightarrow \W^{k_{2},p}(B_{2R}(0))$ ($k_{1} > k_{2}$)
shows that there exist a subsequence $\{\chi_{R}vu_{n_{i}}\}$ such
that 
\leqn{esoblem3.6}{
\chi_{R}vu_{n_{i}} \rightarrow f_{R} \quad \text{strongly in $\W^{k_{2},p}(B_{2R}(0))$}
}
for some $f_{R}$ in $\W^{k_{2},p}(B_{2R}(0))$. Since $\chi_{R} = 1$
on $B_{R}(0)$ we must have that 
\leqn{esoblem3.7}{
f_{R} = vu \quad \text{on $B_{R}(0)$}.
}
Setting $c_{R} = \sup_{r\in [0,R]}e^{-\mu p r}$, we get
\alin{esoblem3.8}{
\nnorm{vu-vu_{n_{i}}}^{p}_{k_{2},p,\mu} & \leq 
c_{R}\norm{vu-vu_{n_{i}}}^{p}_{k_{2},p;B_{R}(0)} + 
 K\xi(R)^{p}\nnorm{u-u_{n_{i}}}^{p}_{k_{2},p,\mu;E_{R}} \\
& \leq c_{R} \norm{vu-vu_{n_{i}}}^{p}_{k_{2},p;B_{R}(0)}
+  K\xi(R)^{p}\nnorm{u-u_{n_{i}}}^{p}_{k_{1},p,\mu} \\
& \leq  c_{R}\norm{vu-vu_{n_{i}}}^{p}_{k_{2},p;B_{R}(0)}
+ 2^{p}K\xi(R)^{p} 
}
where K is a constant independent of R and in getting the last inequality we used
$\nnorm{u_{n}}_{k_{1},p,\mu},\nnorm{u}_{k_{1},p,\mu}\leq 1$. For fixed
$\epsilon > 0$ we can choose $R$ large enough so that 
\eqn{esoblem3.9}{
2^{p}K\xi(R)^{p} \leq \frac{\epsilon^{p}}{2} \, .
}
With $R$ fixed, we get by \eqref{esoblem3.6} and \eqref{esoblem3.7} that there exists
an $M > 0$ such that
\eqn{esoblem3.10}{
c_{R}\norm{vu-vu_{n_{i}}}^{p}_{k_{2},p;B_{R}(0)} \leq
 \frac{\epsilon^{p}}{2} \quad \text{for $i\geq M$} \, .
}
Therefore $\nnorm{vu-vu_{n_{i}}}_{k_{2},p,\mu} \leq \epsilon$ for $i\geq M$
and hence $vu_{n_{i}}$ converges to $vu$ in $\Wc^{k_{2},p}_{\mu}$. This
proves that the map $L_{v}(u)=uv$ is compact.
\end{proof}

The exponentially weighted Sobolev and H\"{o}lder inequalities can 
also be used
to prove a multiplication lemma as in the \rad weighted case (see
lemma \ref{multiply}).
\begin{lem}  \label{emultiply} \mnote{[emultiply]}
If there exists a multiplication
$V_{1}\times V_{2} \rightarrow V_{3}$ $(u,v)\mapsto u\cdot v$ then
for $1\leq p < \infty$
the corresponding multiplication
\eqn{emultiply1}{
\Wc^{k_{1},p}_{\mu_{1}}(\Rbb^{n},V_{1})\times \Wc^{k_{2},p}_{\mu_{2}}(\Rbb^{
n},V_{2})
\rightarrow \Wc^{k_{3},p}_{\mu_{3}}(\Rbb^{n},V_{3})\; : \; (u,v)
\mapsto u\cdot v
}
is bilinear and continuous if
$k_{1},k_{2} \geq k_{3}$, $k_{3} < k_{1} + k_{2} - n/p$, and
$\mu_{1} +\mu_{2} \leq \mu_{3}$ .
\end{lem}

As with the local Sobolev inequaltites, local estimates for elliptic
operators can be extended to global ones on the exponentially weighted
spaces. 
\begin{prop} \label{esobprop4} \mnote{[esobprop4]}
Let $1 < p < \infty$, and $P$ be the elliptic operator defined by
\eqn{esobprop4.1}{
Pu = a^{ij}\partial^{2}_{ij}u + b^{i}(x)\partial_{i}u + c(x)u
} 
where $b^{i},c\in L^{\infty}(\Rbb^{n})$ and there exists constants $\lambda >0$, $0<\alpha\leq 1$
such that $a^{ij}\in C^{0,\alpha}_{0}(\Rbb^{n})$ and 
$\lambda|\xi|^2 \leq a^{ij}(x)\xi_{i}\xi_{j}\leq  \lambda^{-1} |\xi|^2$ for all
for all $x,\xi\in \Rbb^{n}$. Then $P$ defines a continuous map from 
$\Wc^{2,p}_{\mu} \rightarrow \Wc^{0,p}_{\mu}$. Moreover, if $u\in \Wc^{0,p}_{\mu}$
and $Pu\in \Wc^{0,p}_{\mu}$ then $u\in \Wc^{2,p}_{\mu}$ and there exists a constant
$C = C(n,p,\lambda,\norm{a^{ij}}_{C^{0,\alpha}_{0}},\norm{b^{i}}_{\infty},\norm{c}_{\infty})$
such that
\eqn{esobprop4.2}{
\nnorm{u}_{2,p,\mu} \leq C\bigl( \nnorm{Pu}_{0,p,\mu}+\nnorm{u}_{0,p,\mu}
\bigr) \, .
}
\end{prop}
\begin{proof}
If $u\in \Wc^{0,p}_{\mu}$
and $Pu\in \Wc^{0,p}_{\mu}$, then elliptic regularity shows that $u \in \Wc^{2,p}_{\text{loc}}$. The
proof then follows from the local elliptic estimates (see \cite{GilTru98}, theorem 9.11) and the
covering argument.
\end{proof}

In the analysis of elliptic operators on the \rad weighted spaces
the Laplacian $\Delta$ played a fundamental role. The corresponding
fundamental elliptic operator on the exponentially weighted spaces is 
\leqn{fundamental}{
-\Delta + \kappa^2 \quad \text{where $\kappa > 0$ is a constant.}
}
With our applications in mind,  we will restrict ourselves to $n=3$ for
the remainder of this section. The operator \eqref{fundamental}
has a Green's function $G_{\kappa}(x,y)$ which for $n=3$ is
\leqn{Green}{
G_{\kappa}(x,y) = G_{\kappa}(x-y) = \frac{1}{4\pi}
\frac{e^{-\kappa\enorm{x-y}}}{\enorm{x-y}} 
}
and is known as the Yukawa potential. It satisfies the
distributional identity
\leqn{distid}{
(-\Delta_{x} + \kappa^2)G_{\kappa}(x,y) = \delta(x-y) \quad
\text{in $\Dc'(\Rbb^3)$}\, .
}
The invertibility of the operator \eqref{fundamental} can
be established from an estimate for the Green's function
combined with the weighted elliptic estimates in a similar
fashion as for the Laplacian \cite{Bart86}. 
\begin{thm} \label{Yukawa} \mnote{[Yukawa]}
If $\kappa-|\mu| > 0$, $1 < p < \infty$, and $s$ is
a non-negative integer  then the  operator
\leqn{Yukawa.1}{
-\Delta + \kappa^2 : \Wc^{2+s,p}_{\mu} \longrightarrow \Wc^{s,p}_{\mu}
}
is an isomorphism with the inverse
given by
\leqn{Yukawa.2}{
(-\Delta + \kappa^2)^{-1}u(x) = \frac{1}{4\pi}\int_{\Rbb^{3}}
\frac{e^{-\kappa\enorm{x-y}}}{\enorm{x-y}}u(y)dy \, .
}
\end{thm}
\begin{proof}
It suffices to prove the theorem for $s=0$. Let $\Gh_{\kappa}$ be
the operator defined by
\leqn{Yukawa.3}{
\Gh_{\kappa}(u)(x) := \int_{\Rbb^{3}}G_{\kappa}(x-y)u(y)dy \, .
}
\begin{lem} \label{sharp} \mnote{[sharp]}
If $p\geq 1 $, $\kappa - |\mu| > 0$ and $u\in \Lc^{p}_{\mu}$,
then
\eqn{sharp.1}{
\nnorm{\Gh_{\kappa}(u)}_{p,\mu} \leq C\nnorm{u}_{p,\mu}
}
for a constant $C$ independent of $u$.
\end{lem}
\begin{proof}
For all $\mu \in \Rbb$ and $x,y\in \Rbb^{3}$
it holds that $\mu|y|-\mu|x| \leq |\mu||x-y|$
and hence
\eqn{sharp.2}{
e^{\mu|y|-\mu|x|}\leq e^{|\mu||x-y|} \, .
} 
Using this and  the definition of the Green's function
\eqref{Green}, we see that for two non-negative functions $u,v$
\eqn{sharp.3}{
0 \leq u(x)e^{-\mu|x|}G_{\kappa}(x-y)e^{\mu|y|}v(y)
\leq u(x)G_{\kappa-|\mu|}(x-y)v(y) \, .
}
Integrating gives
\lalign{sharp.4a}{
\int_{\Rbb^{3}}\int_{\Rbb^{3}}
 u(x)e^{-\mu|x|}G_{\kappa}(x-y)&e^{\mu|y|}v(y)dxdy \notag \\
& \leq \int_{\Rbb^{3}}\int_{\Rbb^{3}}u(x)G_{\kappa-|\mu|}(x-y)v(y)dxdy 
\label{sharp.4}\, .
} 
Noting that $G_{\kappa-|\mu|}\in L^{1}(\Rbb^{3})$ for
$\kappa - |\mu| > 0$, Young's inequality (see \cite{LL}, theorem 4.2) applied to \eqref{sharp.4}
yields
\leqn{sharp.5}{
\int_{\Rbb^{3}}\int_{\Rbb^{3}}
 u(x)e^{-\mu|x|}G_{\kappa}(x-y)e^{\mu|y|}v(y)dxdy
\leq C\norm{u}_{p'}\norm{G_{\kappa-|\mu|}}_{1}\norm{v}_{p}
}
where $\frac{1}{p}+\frac{1}{p'} = 1$, $p\geq 1$.
Setting
\eqn{sharp.6}{
u(x) = 
\left(\int_{\Rbb^{3}}e^{-\mu|x|}G_{\kappa}(x-y)e^{\mu|y|}v(y)dy\right)^{p/p'}
}
in \eqref{sharp.5} yields
\leqn{sharp.7}{
\left(\int_{\Rbb^{3}} \left| \int_{\Rbb^{3}} 
e^{-\mu|x|}G_{\kappa}(x-y)e^{\mu|y|}v(y)dy
\right|^{p}dx\right)^{1/p}
\leq C\norm{G_{\kappa-|\mu|}}_{1}\norm{v}_{p} \,.
}
Finally, setting $v(y) = e^{-\mu|y|}w(y)$ in \eqref{sharp.7}
shows that
\eqn{sharp.8}{
\nnorm{\Gh_{\kappa}(w)}_{p,\mu} \leq C\norm{G_{\kappa-|\mu|}}_{1}
\nnorm{w}_{p,\mu} \, .
}
So far our above choices amount to assuming that $w\geq 0$. However,
it is clear that the above inequality extends to all $w\in \Lc^{p}_{\mu}$.
\end{proof}
The distributional identity \eqref{distid}
shows that $\Gh_{\kappa}(\Delta u-\kappa^{2}u) = -u$ for all $u\in \Co(\Rbb^{3})$,
and hence 
\eqn{Yukawa.4}{
\nnorm{u}_{0,p,\mu} \leq C\nnorm{\Delta u -\kappa^{2}u}_{0,p,\mu}
\quad \text{for all $u\in  \Wc^{0,p}_{\mu}$}
}
by lemma \ref{sharp} and the density of $\Co(\Rbb^{3})$ in 
$\Wc^{0,p}_{\mu}$. Applying proposition \eqref{esobprop4} to the
above inequality then yields
\leqn{Yukawa.5}{
\nnorm{u}_{2,k,p} \leq C\nnorm{\Delta u -\kappa^{2}u}_{0,p,\mu}
\quad \text{for all $u\in  \Wc^{0,p}_{\mu}$.}
}
Since  $-\Delta +\kappa^{2} : \Wc^{2,p}_{\mu}\rightarrow \Wc^{0,p}_{\mu}$ 
is bounded, it follows easily from \eqref{Yukawa.5} that  $-\Delta +\kappa^{2}$
has closed range and a trivial kernel. The distributional
identity \eqref{distid} implies that $(-\Delta +\kappa^{2})\Gh_{\kappa}(u)
=u $ for all $u\in \Co(\Rbb^{3})$. But by lemma 2.1 $\Gh_{\kappa}(u)\in \Wc^{0,p}_{\mu}$
and hence $\Gh_{\kappa}(u) \in \Wc^{2,p}_{\mu}$ by proposition \ref{esobprop4}. 
Therefore $-\Delta+\kappa^{2}$ is surjective.
\end{proof}

%% file: sss.tex
\sect{sss}{Static spherically symmetric fields}

We assume that all the  fields are static and that $\partial_{0}$
is a timelike hypersurface orthogonal killing vector field for the
metric. Therefore
\eqn{static1}{
\partial_{0} \U^{\alpha \beta} = 0\, , \; \partial_{0} A_{\alpha} = 0\, ,\;
 \partial_{0} \Phi = 0\, 
\quad \text{and}\quad \U^{j 0} = \U^{0 j} = 0\, .
}
Since $\U^{\alpha \beta}$ is symmetric, i.e. $\U^{\alpha \beta} = 
\U^{\beta \alpha}$, we define the following subspace of
the 4 by 4 matrices
\eqn{static3}{
\Sbb := \{\, X=(X^{\alpha \beta}) \in \Mbb_{4\times 4} \,|\, X^{\alpha \beta}
= X^{\beta \alpha} \;\text{and}\; X^{0 j} = 0 \,\}\;.
}
Then letting $\U = (\U^{\alpha\beta})$, $\U$ takes values
in $\Sbb$.

In addition to being static, we will also assume that our fields 
are spherically symmetric.
To define what we mean by spherical symmetry we first
need to specify an action of $SO(3)$ on spacetime $\Rbb^{4}$.
We want $SO(3)$ to act on the hypersurfaces
orthogonal to the timelike Killing vector field $\partial_{0}$.
So using the matrix representation of $SO(3)$ given by
\eqn{sorep}{
SO(3) = \{\, a\in \Mbb_{3\times 3}\,|\, \text{$a^{t} = a^{-1}$ and
$\det(a)=1$} \,\}
}
we define a $SO(3)$ action on spacetime by
\eqn{soaction}{
\rho : SO(3) \times \Rbb^{4} \rightarrow \Rbb^{4} \; :\; (a,(x^{0},x)) \rightarrow
\Phi_{a}(x^{0},x):=(x^{0},ax)
}
where we are treating $x$ as a column vector and $ax$ denotes matrix
multiplication. We then get the induced action on functions via pullbacks.
Lifting the $SO(3)$ action on spacetime to the tensor bundle, we
get the following action on the static metric densities
\eqn{soaction3}{
\rho_{a}(\U)(x) := \at \U(a^{\text{t}}x))\at^{\text{t}} 
}
where
\eqn{soaction4}{
\at := \begin{pmatrix} 1 & 0\\ 0 & a \end{pmatrix}\, .
}
This allows us to define the set of static smooth $SO(3)$-invariant 
metric densities by
\eqn{cssdens}{
\sUo := \{\, \U \in \Co(\Rbb^{3},\Sbb) \, | \,
\U = \rho_{a} \U \; \text{for all}\; a \in SO(3) \, \} \, .
}
Completing in the $\W^{k,p}_{\delta}$ norm yields
\leqn{sspace1}{
\Uc^{k,p}_{\delta}:= \overline{\sUo} \subset
\W^{k,p}_{\delta}(\Rbb^{3},\Sbb) \, .
}
\begin{prop} \label{slaplaceB} \mnote{[slaplaceB]}
For $-1 <\delta < 0$, $1 < p < \infty$ and $k\in \Nbb_{0}$  the Laplacian
$\Delta : \Uc^{k+2,p}_{\delta} \rightarrow
\Uc^{k,p}_{\delta-2}$
is an isomorphism.
\end{prop}
\begin{proof}
From proposition 2.2 of \cite{Bart86}
we have that $\Delta : \W^{k,p}_{\delta}(\Rbb^{3},\Sbb)\rightarrow
\W^{k-2,p}_{\delta-2}(\Rbb^{3},\Sbb)$ is an isomorphism for $1<p<\infty$,
$-1<\delta < 0$. A straightforward calculation shows that 
$\Delta(\sUo) \subset \sUo$.
Similarly, using the formula 
\eqn{laplace2}{
(\Delta^{-1}\U^{\alpha\beta})(x) = 
\frac{-1}{4\pi}\int_{\Rbb^{3}} \frac{\U^{\alpha\beta}(y)}{|x-y|}
d^{3}y\, 
}
it is not difficult to verify that
if $\U \in \sUo$ then $\rho_{a}(\Delta^{-1}\U) = 
\Delta^{-1}\U$ for all $a\in SO(3)$. But $\sUo$ is dense
in $\Uc^{k,p}_{\delta}$ and hence the proof follows.
\end{proof}

Let $\sCo(\Rbb^{3})$ denote the set of smooth $SO(3)$-invariant functions with
compact support, i.e.
\eqn{cssdil}{
\sCo(\Rbb^{3}) := \{\, \phi \in \Co(\Rbb^{3})\, | \, \phi = 
\rho_{a}^{*}\phi \;\text{for all} \; a\in SO(3)\,\} \, .
}
In other words, $\sCo(\Rbb^{3})$ is the set of radial functions
on $\Rbb^{3}$. We then define the space of static spherically symmetric 
Higgs fields with compact support by
\leqn{sspace2}{
\Hc^{\infty}_{0} := \{ \phi(x)x^{j}\tau_{j} \, | \,
\phi \in \sCo(\Rbb^{3})\, \}\, 
}
where
\eqn{su2basis}{
\tau_{1} = \frac{1}{2i}\begin{pmatrix} 0 & 1\\ 1& 0 \end{pmatrix}\, , \;
\tau_{2} = \frac{1}{2i}\begin{pmatrix} 0 & -i\\ i& 0 \end{pmatrix}\, , \;
\tau_{3} = \frac{1}{2i}\begin{pmatrix} 1 & 0\\ 0& -1 \end{pmatrix}\, ,
}
is a basis for $\sU{2}$. We will choose the normalization of the
Ad-invariant inner-product $\ip{\cdot}{\cdot}$ so that
\eqn{normalize}{
\ip{\tau_{i}}{\tau_{j}} = \delta_{ij} \, .
}
Completing $\Hc^{\infty}_{0}$ in the 
$\W^{k,p}_{\delta}(\Rbb^{3},\sU{2})$ norm gives
\eqn{sspace3}{
\Hc^{k,p}_{\delta} := \overline{\Hc^{\infty}_{0}} \subset
\W^{k,p}_{\delta}(\Rbb^{3},\sU{2}) \, .
}

\begin{prop} \label{slaplaceA} \mnote{[slaplaceA]}
Suppose $f \in C^{\infty}([0,\infty))$ satisfies
 $1-f(r)=\Ord(r^{2})$ as $r\rightarrow 0$, $f(r) = \Ord(r^{-\eta})$
as $r\rightarrow \infty$ for some $\eta > 0$,
and $f \geq 0$. Then
for $1<p<\infty$, $-1<\delta < 0$, and $k\in \Nbb_{0}$ the operator
\eqn{slaplaceA.1}{
\Hc^{k+2,p}_{\delta} \longrightarrow 
\Hc^{k,p}_{\delta-2}\; : \; \Phi(x) \longmapsto  \Delta \Phi(x) + 
\frac{2}{|x|^{2}}(1-f(|x|))\Phi(x)
}
is an isomorphism.
\end{prop}
\begin{proof}
Without loss of generality we can assume that $k=0$.
We first show that the
operator
\leqn{slaplaceA.2}{
P := \Delta + 2|x|^{-2}(1-f(|x|))
}
has a finite dimensional kernel and closed range on the space of
static spherically symmetric Higgs fields.
\begin{lem} \label{slaplaceAlem1} \mnote{[slaplaceAlem1]}
For $-1<\delta<0$, $1<p<\infty$ the operator $P$ defines a continuous map from
$\Hc^{2,p}_{\delta} \rightarrow \Hc^{0,p}_{\delta-2}$ that
has closed range and a finite dimensional kernel.
\end{lem}
\begin{proof}  
Directly from the definition of the weighted spaces it is easy to see that
$P$
defines a continuous map from $\W^{2,p}_{\delta}(\Rbb^{3},\sU{2})$ to
$\W^{0,p}_{\delta-2}(\Rbb^{3},\sU{2})$. A calculation shows
that $P(\Hc^{\infty}_{0}) \subset \Hc^{\infty}_{0}$ and hence
$P$ defines a continuous map from 
$\Hc^{2,p}_{\delta} \rightarrow \Hc^{0,p}_{\delta-2}$. 

Suppose
$\Phi \in \Hc^{\infty}_{0}$. Then split $\Phi$ as 
$\Phi = \Phi_{0} + \Phi_{\infty}$ 
where $\Phi_{0} = \chi_{2} \Phi$ and $\Phi_{\infty} = (1-\chi_{2})\Phi$. Since
$\Phi(x) = \phi(x)x^{j}\tau_{j}$ for some $\phi\in \sCo(\Rbb^{3})$,
\eqn{slaplaceA.4}{
\Phi_{\infty}(x) = \frac{\phi_{\infty}(x)}{|x|}x^{j}\tau_{j}
}
where $\phi_{\infty}(x) := |x|(1-\chi_{2}(x))\phi(x) \in \sCo(\Rbb^{3})$. 

Straightforward calculation verifies that  for $|x| > 0$,
$|\Phi_{\infty}|^{2} = |\phi_{\infty}|^{2}$, 
$|\partial \Phi_{\infty}|^{2} = |\partial \phi_{\infty}|^{2} + 2|x|^{-2}|\phi_{\infty}|^{2}$
and $|\partial^{2}\Phi_{\infty}|^{2} = |\partial^{2}\phi_{\infty}|^{2} + 8|x|^{2-}|\partial \phi_{\infty}|^{2}
+  6|x|^{-4}|\phi_{\infty}|^{2}$.
Using this and $\supp |\phi|,\supp |\Phi| \subset 
E_{1}(0)$
 it follows that there exists a $C$ independent of $\phi$
and $\Phi$ such that
\leqn{slaplaceA.6}{
\norm{\Phi_{\infty}}_{2,p,\delta} \leq C \norm{\phi_{\infty}}_{2,p,\delta}\,.
}
A short calculation shows that
\leqn{slaplaceA.7}{
P\Phi_{\infty}(x) = \Bigl(\Delta\phi_{\infty}(x) - 
\frac{2f(|x|)}{|x|^2}\phi_{\infty}(x)\Bigr)\frac{x^{k}}{|x|}\tau_{k}\,.
}
Thus if we define 
\eqn{slaplaceA.8}{
Q := \Delta -\frac{2(1-\chi_{1/4}(x))f(|x|)}{|x|^2} \, ,
}
then $P\Phi_{\infty}(x) = Q\phi_{\infty}(x)\frac{x^{k}}{|x|}\tau_{k}$
since $\supp |\phi_{\infty}|\subset E_{0}$. So
$|P\Phi_{\infty}(x)| = |Q\phi_{\infty}(x)|$ and hence
\leqn{slaplaceA.9}{
\norm{P\Phi_{\infty}}_{0,p,\delta-2} = 
\norm{Q\phi_{\infty}}_{0,p,\delta-2} \, .
}
In the terminology of \cite{Bart86}, the operator $Q$ is
asymptotic to $\Delta$. Therefore by \cite{Bart86} theorem 1.10 we
have the estimate
\leqn{slaplaceA.10}{
\norm{\phi_{\infty}}_{2,p,\delta} 
\leq C \bigl(\norm{Q\phi_{\infty}}_{0,p,\delta-2}
+ \norm{\phi_{\infty}}_{p;B_{R}(0)}\bigl)
}
for some $R > 0$. Since $ \norm{\phi_{\infty}}_{p;B_{R}(0)} =
 \norm{\Phi_{\infty}}_{p;B_{R}(0)}$, we get the following
estimate from \eqref{slaplaceA.6}, \eqref{slaplaceA.9}, and
\eqref{slaplaceA.10}
\leqn{slaplaceA.11}{
\norm{\Phi_{\infty}}_{2,p,\delta}
\leq C \bigl(\norm{P\Phi_{\infty}}_{0,p,\delta-2}
+ \norm{\Phi_{\infty}}_{p;B_{R}(0)}\bigl) \, .
}
Once we have this scale broken estimate we can proceed as
in the proof of theorem 1.10 of \cite{Bart86} to conclude that
$P$ has closed range and a finite dimensional kernel.
\end{proof}
With respect to the pairing $(\Psi,\Phi) = \int \ip{\Psi}{\Phi}d^{3}x$
the operator has a formal adjoint $P^{*}=P$. Since 
$W^{0,p}_{\delta-2}(\Rbb^{3},\sU{2})^{*}=W^{0,p'}_{-1-\delta}(\Rbb^{3},\sU{2})$
where $p'=p/(p-1)$, it follows from 
proposition \ref{eestimate} and proposition 1.14 of \cite{Bart86} that
$\text{ker}{P^{*}}\subset W^{2,p}_{-1-\delta}(\Rbb^{3},\sU{2})$.
Therefore by the above lemma
\leqn{slaplaceA.11a}{
\dim\text{coker} P\bigl|_{\Hc^{2,p}_{\delta}} = 
\dim \text{ker} P\bigl|_{\Hc^{2,p}_{-1-\delta}} < \infty .  
}

\begin{lem} \label{tker} \mnote{[tker]}
For any $\delta < 0$, $1<p<\infty$, $\ker P\bigl|_{\Hc^{2,p}_{\delta}}
= \{0\}$ .
\end{lem}
\begin{proof}
Suppose $\Psi \in \Hc^{2,p}_{\delta}$ satisfies $P\Psi = 0$. Then
by elliptic regularity (see \cite{GilTru98} theorem 9.19
or \cite{Giaq93} theorem 3.6),
$P\Psi=0$ implies that $\Psi \in C^{\infty}(\Rbb^{3},\sU{2})$.
So there exists a function $\psi(r) \in C^{\infty}([0,\infty))$ such that
\eqn{slaplaceA.12}{
\Psi(x) = \psi(|x|)\frac{x^{j}}{|x|}\tau_{j} \quad
\text{and}\quad \psi(r) = cr + \Ord(r^3) \quad \text{as $r\rightarrow 0$} \, .
}
It follows from the
equality $|\Psi(x)|=|\psi(|x|)|$ and 
theorems 1.2 of \cite{Bart86} that
$\psi(x) = \ord(|x|^{\delta})$,
$\partial_{i}\psi(x) = \ord(|x|^{\delta-1})$, and
$\partial_{i}\partial_{j}\psi(x)=\ord(|x|^{\delta-2})$
as $|x|\rightarrow \infty$.
Now $P\Psi=0$ implies that (see \eqref{slaplaceA.7}) 
$\Delta \psi(x) - |x|^{-2}2f(|x|)\psi(x)=0$.
Multiplying by $\psi(x)$ yields 
$\psi\Delta\psi - 2f(|x|)|x|^{-2}\psi^2 = 0$ which by
the fall off conditions for $\psi$ near $|x|=0$ and
$|x|=\infty$ is integrable. 
Integrating yields
\eqn{slaplaceA.15}{
\int_{\Rbb^{3}}\psi\Delta\psi d^{3}x - 
\int_{\Rbb^{3}} \frac{2f(|x|)}{|x|^2}\psi^2 d^{3}x = 0 \, .  
}
Integrating by parts which is again valid by the fall off conditions
conditions then gives 
\eqn{slapalceA.16}{
\int_{\Rbb^{3}}|\partial\psi|^2 d^{3}x +
\int_{\Rbb^{3}}\frac{2f(|x|)}{|x|^2}\psi^2 d^{3}x = 0 \, .
}
Thus $f\geq 0$ implies that $\psi=0$ and hence 
$\Psi=0$. 
\end{proof}
The proof now follows from \eqref{slaplaceA.11a} and lemmas
\ref{slaplaceAlem1} and \ref{tker} which imply that
$\dim\text{coker} P\bigl|_{\Hc^{2,p}_{\delta}} = 
\dim \ker  P\bigl|_{\Hc^{2,p}_{\delta}} = 0$.
\end{proof}

In addition to being spherically symmetric, we will
assume that our gauge potential is purely magnetic.
Choosing an appropriate gauge, the gauge potential
can then be written as \cite{k5109}
\eqn{sgpot}{
A_{0} = 0 \quad \text{and} \quad A_{i}(x) := a(|x|)\epsilon_{i}{}^{j}{}_{k} x^{k}\tau_{j} \, .
}
If we write the gauge potential $A_{i}$ as a 3-tuple
$A = (A_{1},A_{2},A_{3})$ then the gauge potential
$A$ takes values in the space $\sU{2}^{3}$ which
carries a norm
\eqn{gpotnorm}{
\enorm{A}^{2} := \sum_{i=1}^{3} \ip{A_{i}}{A_{i}} \, .
}

We define the  set of smooth static spherically symmetric 
purely magnetic gauge potentials with compact support
by
\eqn{cssgpot}{
\sAo := \{ A : \Rbb^{3} \rightarrow 
\sU{2}^{3}\, | \, A_{i}(x) = 
a(x)\epsilon_{i}{}^{j}{}_{k} x^{k}\tau_{j}\; \text{for some} \;a\in \sCo(\Rbb^{3})
\, \} \, .
}
Completing this in the $\Wc^{k,p}_{\mu}(\Rbb^{3},\sU{2}^{3})$
gives
\eqn{sspace4}{
\Ac^{k,p}_{\mu} = \overline{\sAo} \subset \Wc^{k,p}_{\mu}(\Rbb^{3},\sU{2}^{3})\,.
}
Notice that every $A \in \sAo$ satisfies
\eqn{divA1}{
\Div A := \sum_{j=1}^{3} \partial_{j} A_{j} = 0 \, ,
}
which implies by the continuity of differentiation
(see \eqref{ediff}) that
\leqn{divA2}{
\Div A = 0 \quad \text{for all $A \in \Ac^{k,p}_{\mu}$}.
}
This shows that the static spherically symmetric gauge potentials satisfy the 
Coulomb gauge condition globally on $\Rbb^{3}$. As is well known, this is
a very special situation and is one of the reasons that makes 
the static spherically symmetric Yang-Mills equations easy to analyze.

\begin{prop} \label{sYukawa} \mnote{[sYukawa]}
For $\kappa -|\mu|> 0$, $k\in \Nbb_{0}$ and $1<p<\infty$,
the operator $\Delta-\kappa^2 : \Ac^{k+2,p}_{\mu} \rightarrow
\Ac^{k,p}_{\mu}$ is an isomorphism.
\end{prop}
\begin{proof}
Follows from directly from theorem \ref{Yukawa} using the
same arguments as in the proof of proposition \ref{slaplaceB}.
\end{proof}

%% file: mymh.tex
\sect{mymh}{The modified Yang-Mills equation}

Instead of solving the Yang-Mills equation \eqref{YMHa1}
we will instead solve a related system of equations whose solutions
will also be solutions to \eqref{YMHa1}. 
The reason for this modification is to make the Yang-Mills equation  
differentiable on the spherically symmetric function spaces introduced
in section \ref{sss}.  

We begin by splitting the YM potential and the Higgs fields. Let 
\leqn{yh1}{
Y = Y_{j}dx^{j} = \frac{\chi_{1}(x)-1}{r^2}\epsilon_{i}{}^{j}{}_{k}x^{k}\tau_{j}
\quad , \quad \Omega = \frac{1-\chi_{1}(x)}{r}x^{j}\tau_{j} \, .
}
and
\eqn{yh2}{
A = Y+Z , \quad \Phi = \Omega + \Psi 
}
where $Z\in \Ac^{k,p}_{\mu}$ and $\Psi\in \Hc^{k,p}_{\delta}$ 
will be considered as the unknowns. 
Assume for the moment
that $Z$ and $\Psi$ are $C^{1}$ and spherically symmetric. Then we can write 
\eqn{yh3}{
Z_{j} = z(r)\epsilon_{i}{}^{j}{}_{k}x^{k}\tau_{j}
\quad \text{and} \quad \Psi = \psi(r)x^{j}\tau_{j}
}
and a short calculation shows that
$[\Phi,D_{0}\Phi] = 0$ and
\eqn{yh4}{
[\Phi,D_{i}\Phi] = \Bigl(\frac{1-\chi_{1}}{r}+\psi\Bigr)^{2}
\bigl(\chi_{1}+ r^{2} z\bigr)\epsilon_{i}{}^{j}{}_{k}x^{k}\tau_{j}
= |\Omega+\Psi|^{2}\Bigl(\frac{\chi_{1}}{r^{2}}
\epsilon_{i}{}^{j}{}_{k}x^{k}\tau_{j}+ Z_{i}\Bigr)\, .
}
Thus for $C^{1}$ static spherically symmetric fields we have the identity
\leqn{yh5}{
[\Phi,D_{\alpha}\Phi] =(1-\delta^{0}_{\alpha})\Bigl( \chi_{1}[\Phi,D_{\alpha}\Phi]+(1-\chi_{1})
|\Phi|^{2}\Bigl(\frac{\chi_{1}}{r^{2}}
\delta^{i}_{\alpha}\epsilon_{i}{}^{j}{}_{k}x^{k}\tau_{j}+ \delta^i_\alpha
Z_{i}\Bigr)\Bigr)\, .
}
This motivates us to consider the following modified Yang-Mills
equation 
\lgath{yh6ab}{
\gb^{\alpha\nu}\bigl((1-\chi_{3})D^{Y}_{\alpha}F^{Y}_{\nu\beta}
+D^{A}_{\alpha}F^{A}_{\nu\beta}\bigr) \notag \\
-(1-\delta^{0}_{\beta})\Bigl(
\chi_{1}[\Phi,D_{\beta}\Phi] +(1-\chi_{1})
|\Omega+\Psi|^{2}\Bigl(\frac{\chi_{1}}{r^{2}}
\delta_{\beta}^{i}\epsilon_{i}{}^{j}{}_{k}x^{k}\tau_{j}+ Z_{\beta}\Bigr)
\Bigr) = 0 \label{yh6a}
}
where $A=Y+Z$ and $\Phi=\Psi+\Omega$.
Observe that if the term $(1-\chi_{3})D^{Y}_{\alpha}F^{Y}_{\nu\beta}$
vanished then this equation would be the same as equation \eqref{YMHa1} modified
by the identity \eqref{yh5} and written in term of the new variables $Z$ and $\Psi$. 
We shall see later that for static spherically
symmetric solutions the $(1-\chi_{3})D^{Y}_{\alpha}F^{Y}_{\nu\beta}$ does
vanish.  This will show that solutions to \eqref{yh6a} will be solutions
to \eqref{YMHa1}.  Our assumption that the fields are static
and spherically symmetric imply that $F^{A}_{\alpha 0} = 0$, $F^{Y}_{\alpha 0} =0$
and $\Gamma^{k}_{l0}=0$ and hence equation \eqref{yh6a} will be satisfied automatically
for $\beta = 0$. Therefore we need only solve
\lgath{yh6b}{
\gb^{\alpha\nu}\bigl((1-\chi_{3})D^{Y}_{\alpha}F^{Y}_{\nu i}
+D^{A}_{\alpha}F^{A}_{\nu i}\bigr)
- \notag \\
\Bigl(\chi_{1}[\Phi,D_{i}\Phi] + (1-\chi_{1})
|\Phi|^{2}\Bigl(\frac{\chi_{1}}{r^{2}}
\epsilon_{i}{}^{j}{}_{k}x^{k}\tau_{j}+ Z_{i}\Bigr)
\Bigr) = 0 \label{yh6b.1} \, . 
}

In terms of the new variables $Z$ and $\Psi$ the Higgs equations \eqref{YMHa2}
becomes
\lgath{yh7}{
\gb^{\alpha\beta}\Bigl( \partial_{\alpha}D^{Y}_{\beta}\Omega - \Gamma_{\alpha\beta}^{\sigma}
D^{Y}_{\sigma}\Omega +[Y_{\alpha},D^{Y}_{\beta}\Omega] + 
\partial_{\alpha}D^{Y}_{\beta}\Psi - \Gamma_{\alpha\beta}^{\sigma}
D^{Y}_{\sigma}\Psi +[Y_{\alpha},D^{Y}_{\beta}\Psi] + \notag \\
\partial_{\alpha}[Z_{\beta},\Omega+\Psi] 
-\Gamma_{\alpha\beta}^{\sigma}[Z_{\sigma},\Omega+\Psi]
+[Y_{\alpha},[Z_{\beta},\Omega+\Psi]]+ \notag\\  
[Z_{\alpha},D^{Y}_{\beta}\Omega +
D^{Y}_{\beta}\Psi + [Z_{\beta},\Omega+\Psi]]\Bigr)=0 \, .\label{yh7.1}
}

%% file: smooth.tex
\sect{smooth}{Analyticity of the field equations}

In this section we establish that the reduced
field equations \eqref{reduced3} and the modified YMH equations \eqref{yh6b.1}-\eqref{yh7.1}
define
analytic maps.  For a definition of analytic maps between Banach spaces see
\cite{Deim} definition 15.1. As is standard we will use $\Cw$ to denote the
class of analytic maps. To establish analyticity we will repeatedly 
use the following:  continuous linear and bilinear maps
between Banach spaces are analytic, and the composition of two
analytic maps is again analytic. Also useful is proposition 3.6 of
\cite{Heil95} which shows how analytic functions on $\Rbb$ can
be used to define analytic maps on Banach algebras. 
 
To begin we first fix some
notation. If $V$ is a Banach space with norm $\norm{\cdot}$
then we define $B_{V}(x;R)$ to be the ball of radius $R$
centered at $x\in V$. 
We recall the following results from  \cite{Heil95} which are 
fundamental in establishing analyticity.
\begin{prop}{\emph{[Proposition 3.10,\cite{Heil95}]}}
 \label{HeilA}\mnote{[HeilA]}
Suppose $3/2<p<\infty$ and $-1<\delta <0$.
Then for any $R >0$ there exists a $\Lambda > 0$ such
that the following maps are of class $\Cw$:
\gath{HeilA1}{
(-\Lambda,\Lambda) \times B_{\W^{2,p}_{\delta}(\Rbb^{3},\Sbb)}(0;R)
\rightarrow \W^{2,p}_{\delta}(\Rbb^{3},\Sbb) \; :\;
(\lambda,\U) \mapsto (\gb^{\alpha\beta} - \underset{o}{\gb}{}^{\alpha\beta}) \\
(-\Lambda,\Lambda) \times B_{\W^{k,p}_{\delta}(\Rbb^{3},\Sbb)}(0;R)
\rightarrow \W^{2,p}_{\delta}(\Rbb^{3},\Sbb) \; :\;
(\lambda,\U) \mapsto (\gb_{\alpha\beta} - \underset{o}{\gb}{}_{\alpha\beta}) \\
\intertext{and}
(-\Lambda,\Lambda) \times B_{\W^{2,p}_{\delta}(\Rbb^{3},\Sbb)}(0;R)
\rightarrow \W^{2,p}_{\delta}(\Rbb^{3}) \; :\;
(\lambda,\U) \mapsto |\df|^{q/2}-1
}
for $q=-3,-2,-1,1,2$. Moreover, the following expansions
are valid
\gath{HeilA2}{
|\df|-1 = -4\lambda \U^{00} + \Ord(\lambda^{2}) \, ,\quad
\sqrt{\df}-1 = -2\lambda \U^{00} + \Ord(\lambda^{2}) \, ,\\
\frac{1}{\sqrt{\df}}-1 = 2\lambda \U^{00} + \Ord(\lambda^{2}) \, , \quad
(\gb_{\alpha\beta} - \underset{o}{\gb}{}_{\alpha\beta}) = -4\lambda
(\delta^{0}_{\alpha}\delta^{0}_{\beta})\U^{00} + \Ord(\lambda^{2}) \, .
}
\end{prop} 

\begin{prop}{\emph{[Proposition 6.2,\cite{Heil95}]}}
\label{HeilB}\mnote{[HeilB]}
Suppose $p > 3$ and $-1 < \delta < 0$. Then for any $R>0$ there
exists a $\Lambda >0$ such that the Christofel symbols
\eqn{HeilB1}{
\Gamma^{\alpha}_{\beta\gamma} :
(-\Lambda,\Lambda) \times B_{\W^{2,p}_{\delta}(\Rbb^{3},\Sbb)}(0;R)
\rightarrow \W^{1,p}_{\delta-1}(\Rbb^{3})
}
are of class $\Cw$ for all $\alpha,\beta,\gamma = 0,1,2,3$. Moreover,
the following expansion is valid
\eqn{HeilB2}{
\Gamma^{\alpha}_{\beta\gamma} = \Gamma^{\alpha}_{\beta\gamma}\big|_{\lambda = 0}
+ \Ord(\lambda)
}
where
\eqn{HeilB3}{
\Gamma^{\alpha}_{\beta\gamma}\big|_{\lambda = 0} = \left\{
\begin{array}{ll} \U^{00}{}_{,\alpha} & \text{if $\beta=\gamma=0$ and
$\alpha \neq 0$}\\
0 & \text{otherwise}
\end{array} \right. \, .
}
\end{prop}

It is important to note that
\leqn{gflatupA}{
(\underset{o}{\gb}{}^{\alpha\beta}) = \begin{pmatrix}
-\lambda & 0 & 0 & 0 \\
 0   & 1 & 0 & 0\\
 0 & 0 & 1 & 0 \\
 0 & 0 & 0 & 1
\end{pmatrix}
}
so that
\leqn{gflatupB}{
(\underset{o}{\gb}{}^{\alpha\beta})\big|_{\lambda=0} = \begin{pmatrix}
 0 &  0 & 0 & 0 \\
 0   & 1 & 0 & 0\\
 0 & 0 & 1 & 0 \\
 0 & 0 & 0 & 1
\end{pmatrix}
}
Using the above propositions and the results from section \ref{wsob}
we can establish that the stress-energy tensor
defines an analytic map.
\begin{prop} \label{smoothB}\mnote{[smoothB]}
Suppose $p > 3$, $-1<\delta <0$ and $\mu < 0$.
Then for any $R >0$ there exists a $\Lambda > 0$ such that
\gath{smoothB1}{
T : (-\Lambda,\Lambda) \times
B_{\Uc^{2,p}_{\delta}}(0;R) \times
\Hc^{2,p}_{\delta}\times
\Ac^{2,p}_{\mu}
\longrightarrow \W^{1,p}_{\delta-2}(\Rbb^{3},\Sbb)
\; :\;
(\lambda,\U,\Psi,Z) \longmapsto (T^{\alpha\beta})
}
and
\gath{smoothB2}{
\Tc : (-\Lambda,\Lambda) \times
B_{\Uc^{2,p}_{\delta}}(0;R) \times
\Hc^{2,p}_{\delta}\times
\Ac^{2,p}_{\mu}
\longrightarrow \W^{1,p}_{\delta-2 }(\Rbb^{3},\Sbb)
\; :\;
(\lambda,\U,\Phi,Z) \longmapsto (\Tc^{\alpha\beta})
}
are of class $\Cw$. Moreover, the following expansion is valid
\eqn{smoothB3}{
\Tc^{0\alpha} = \Tc^{\alpha 0 } = \Ord(\lambda) \quad \text{and} \quad
\Tc^{ij} = 4\pi G\To{}^{ij}(\Psi,Z)  + \Ord( \lambda)
}
where
\lalign{smoothB4}{
\To{}^{ij}(\Psi,Z) & =  
\left(\delta^{ik}\delta^{jl}
\ip{\Do{}^{A}_{k}\Phi}{\Do{}^{A}_{l}\Phi} - \Half \delta^{ij}\delta^{kl}
\ip{\Do{}^{A}_{k}\Phi}{\Do{}^{A}_{l}\Phi}
\right)
+ \notag \\
& \left( \delta^{ik}\delta^{jl}
\delta^{mn}
\ip{F^{A}_{km}}{F^{A}_{ln}}-\Quarter \delta^{kl}\delta^{mn}\delta^{ij}
\ip{F^{A}_{km}}{F^{A}_{ln}} \right)\, , \notag
}
and $A=Y+Z$ and $\Phi=\Omega+\Psi$.
\end{prop}
\begin{proof}
Letting $A=Y+Z$, we can write
\leqn{Fsplit}{
F^{A}_{\alpha\beta} = F^{Y}_{\alpha\beta} +F^{Z}_{\alpha\beta} + [Y_{\alpha},Z_{\beta}] + 
[Z_{\alpha},Y_{\beta} ]
}
where
\leqn{FY1}{
F^{Y}_{0\alpha}=0\, , \quad F^{Y}_{ij} = \epsilon_{ijk}
\left[\frac{\chi_{1}'(r)}{r}\left(\delta^{kl}-\frac{x^{k}x^{l}}{r^2}\right)
 + \frac{\chi_{1}(r)^2-1}{r^4}  x^{k}x^{l} \right]\tau_{l}
}
and $(\cdot)' = \frac{d\,}{dr}(\cdot)$. Since $-1<\delta < 0$, we get from \eqref{FY1}
that 
\leqn{FY2}{
F^{Y}_{\alpha\beta} \in \W^{2,p}_{\delta-1}(\Rbb^{3}) \; .
} 
From the definition of $Y$ it is clear that 
\leqn{Y}{
Y\in \Ac^{0,\infty}_{-1} \, .
}
Then since $\mu <0$ and $p > 3$, it follows from  the inclusion \eqref{mudel}, the multiplication lemma
\ref{multiply}, the weighted H\"{o}lder inequality (theorem 1.2 (ii),\cite{Bart86}),
and equations \eqref{Fsplit}, \eqref{FY2}, \eqref{Y} that the map
\leqn{Fmap}{
\Ac^{2,p}_{\mu} \longrightarrow \W^{1,p}_{\delta-1} : Z \longmapsto  F^{A}\; \text{  is analytic.}
}
Also note that for $\Psi \in \W^{k,p}$,  \eqref{Y} implies via the weighted H\"{o}lder inequality
 (theorem 1.2 (ii),\cite{Bart86}) that $[Y_{\alpha},\Psi] \in \W^{2,p}_{\delta-1}$.
There for the map
\leqn{DY}{
D^{Y}_{\alpha} : \W^{2,p}_{\delta}(\Rbb^{3},\sU{2}) \longrightarrow \W^{1,p}_{\delta-1}(\Rbb^{3}) :
\Psi \longmapsto \partial_{\alpha}\Psi + [Y_{\alpha},\Psi]
}
is continuous.
A short calculation shows that
\lalign{Om}{
D^{Y}_{0}\Omega & = 0 \, ,\label{Om1} \\
D^{Y}_{i}\Omega & = \Bigl(-\frac{\chi_{1}'(r)}{r^2}-\frac{(1-\chi_{1}(r))}{r^3}
\chi_{1}(r)\Bigr)x^{i}x^{k}\tau_{k}
+\frac{1-\chi_{1}(r)}{r}\chi_{1}(r)\tau_{i}\,  , \label{Om2} 
}
and
\leqn{DPhi1}{
D^{A}_{\alpha}\Phi = D^{Y}_{\alpha}\Omega + D^{Y}_{\alpha}\Psi + [Z_{\alpha},\Omega+\Psi]
}
Again, because $\mu <0$ and $p > 3$, the inclusion \eqref{mudel}, the multiplication lemma
\ref{multiply} and \eqref{DY}, \eqref{Om1}, \eqref{Om2}, \eqref{DPhi1} imply that
\leqn{Dhi2}{
 \W^{2,p}_{\delta}(\Rbb^{3},\sU{2})\times \Ac^{2,p}_{\mu} \longrightarrow
 \W^{1,p}_{\delta-1}(\Rbb^{3}) : (\Psi,Z) \longmapsto D^{A}_{\alpha}\Phi\; 
}
is analytic. 
The analyticity of the maps now follows from lemma \ref{multiply}, proposition \ref{HeilA},
\eqref{Fmap}, and \eqref{Dhi2}.
\end{proof}

Letting
\lalign{ximap}{
&\Xia^{\alpha\beta}  = E^{\alpha\beta}-\Tc^{\alpha\beta}\, \label{ximap.1} \\
&\Xib  = \gb^{\alpha\beta}D^{A}_{\alpha}D^{A}_{\beta}\Phi \, \label{ximap.2} \\
&\Xic_{i}  = \gb^{\alpha\nu}\bigl(\chi_{3}-1)D^{Y}_{\alpha}F^{Y}_{\nu i}
+D^{A}_{\alpha}F^{A}_{\nu i}\bigr)
- \Bigl(\chi_{1}[\Phi,D_{j}\Phi] + \notag \\
&\qquad \qquad (1-\chi_{1})
|\Phi|^{2}\Bigl(\frac{\chi_{1}}{r^{2}}
\epsilon_{i}{}^{j}{}_{k}x^{k}\tau_{j}+ Z_{i}\Bigr)
\Bigr)  \label{ximap.3}
}
and
\leqn{ximap2}{
\Xi = \bigl(\Xia,\Xib,\Xic\bigl) = \bigl( (\Xia^{\alpha\beta}), \Xib, (\Xic_{i}) \bigl) \, .
}
we collect our field equations \eqref{reduced3}, \eqref{yh6b.1}, and \eqref{yh7.1}
into a single expression
\leqn{ximap3}{
\Xi = 0 \, .
}
\begin{prop} \label{smoothC}\mnote{[smoothC]}
Suppose $p > 3$, $-1<\delta <-3/p$ and $\mu < 0$
and
\eqn{smoothC1a}{
 X = \W^{0,p}_{\delta-2}(\Rbb^{3},\Sbb)
\times \W^{0,p}_{\delta-2}(\Rbb^{3},\sU{2})
\times \Wc^{0,p}_{\mu}(\Rbb^{3},\sU{2}^{3}).
}  
Then for any $R >0$ there exists a $\Lambda > 0$ such that
\eqn{smoothC1}{
\Xi : (-\Lambda,\Lambda) \times
B_{\U^{2,p}_{\delta}}(0;R) \times
\Hc^{2,p}_{\delta}\times
\Ac^{2,p}_{\mu} \longrightarrow X
\; :\;
(\lambda,\U,\Psi,Z) \longmapsto \bigl( (\Xia^{\alpha\beta}), \Xib, (\Xic_{j}) \bigl) 
}
is of class $\Cw$. Moreover the following expansions are valid
\alin{smoothC2}{
&\Xia^{0\alpha}  = \Delta \U^{0\alpha} + \Ord(\lambda)\, ,\\
&\Xia^{ij}  = \Delta \U^{ij} - \delta^{ik}\delta^{jl}\partial_{k}\U^{00}\partial_{l}\U^{00} +
\Half \delta^{ij}\delta^{kl} 
\partial_{k}\U^{00}\partial_{l}\U^{00} - 4\pi G \To{}^{ij} +\Ord(\lambda) \, ,\\
&\Xib  = \delta^{ij}\Do{}^{A}_{i}\Do{}^{A}_{j}\Phi + \Ord(\lambda) \, , \\
&\Xic_{i}  = \delta^{kl}\Do{}^{A}_{k}F^{A}_{li} -\Bigl(\chi_{1}[\Phi,\Do{}^{A}_{i}\Phi]
+(1-\chi_{1})|\Phi|^2\Bigl(\frac{\chi_{1}}{r^{2}}
\epsilon_{i}{}^{j}{}_{k}x^{k}\tau_{j}+ Z_{i}\Bigr)\Bigr) + \Ord(\lambda) \, .
}
where $A=Y+Z$ and $\Phi = \Omega + \Psi$.
\end{prop}
\begin{proof}
This proposition can be proved in a similar manner to the proof of proposition
\ref{smoothB} by using the inclusions \eqref{rincl}, \eqref{reinc}, and
\eqref{mudel}, the two multiplication lemmas \ref{multiply} and \ref{emultiply},
theorem \ref{esobthm1} and \cite{Bart86} theorem 1.2, 
and propositions \ref{HeilA}, \ref{HeilB}, and \ref{smoothB}. Note that
that formulas used in the proof of proposition \ref{smoothB} are also useful. 

The expansion in $\lambda$ can be inferred from \eqref{gflatupA}
and \eqref{gflatupB}, the expansions
in propositions \ref{HeilA}, \ref{HeilB}, and \ref{smoothB},
and 
\eqn{smoothC3}{
\Bigl[(1-\chi_{3})\gb^{\alpha\nu}D^{Y}_{\alpha}F^{Y}_{\nu i}\Bigr]_{\lambda=0}
= (1-\chi_{3})\delta^{jk}\Do{}^{Y}_{j}F^{Y}_{ki} = 0 \,.
}
The last equality can be seen from
\alin{smoothC4}{
\delta^{jk}\Do{}^{Y}_{j}F^{Y}_{ki}dx^{i}
&=\Bigl(\chi_{1}''-\frac{(\chi_{1}^2-1)\chi_{1}}{r^2}\Bigr) 
\bigl(-\sin\phi\,\tau_{1}+\cos\phi\, \tau_{2}\bigr)d\theta \\
& \quad +\Bigl(\chi_{1}''-\frac{(\chi_{1}^2-1)\chi_{1}}{r^2}\Bigr)
\bigl(\tau_{3}-\cot\theta(\sin\phi\,\tau_{2}+\cos\phi\,\tau_{1})\bigr)d\phi
\,
}
where $(\cdot)'=\frac{d\,}{dr}(\cdot)$.
\end{proof}

\begin{prop} \label{smoothD}\mnote{[smoothD]}
Suppose $p > 3$, $-1<\delta <-3/p$ and $\mu < 0$.
Then for any $R >0$ there exists a $\Lambda > 0$ such that
\gath{smoothD1}{
\Xi : (-\Lambda,\Lambda) \times
B_{\U^{2,p}_{\delta}}(0;R) \times
\Hc^{2,p}_{\delta}\times
\Ac^{2,p}_{\mu} \longrightarrow \Uc^{0,p}_{\delta-2}\times \Hc^{0,p}_{\delta-2}\times
\Ac^{0,p}_{\mu}\\
\; :\;
(\lambda,\U,\Psi,Z) \longmapsto \bigl( (\Xia^{\alpha\beta}), \Xib, (\Xic_{j}) \bigl) 
}
is of class $\Cw$. 
\end{prop}
\begin{proof}
For fixed $R$ let $\Lambda$ be as given by proposition \ref{smoothC}. Then it
can be shown by straightforward calculation that 
$\lambda \in (-\Lambda,\Lambda)$, $\U \in \sUo\cap B_{\Uc^{2,p}}(0;R)$, $\Psi \in \sHo \cap \Hc^{2,p}_{\delta}$,
and $Z\in \sAo\cap \Ac^{2,p}_{\mu}$ implies that 
$\Xi(\lambda,\U,\Psi,Z) \in \Uc^{\infty}\times \Hc^{\infty}\times \Ac^{\infty}$.
The result now follows from the continuity of the map $\Xi$ (see proposition \ref{smoothC})
and the density of  $\sUo$, $\sHo$, and $\sAo$.
\end{proof}

%% file: solve.tex
\sect{solve}{Solving the reduced/modified EYMH equations}

We now employ the same method as in \cite{Heil95} to find solutions to
the reduced/modified EYMH equations. Namely, we first solve the reduced
equations for $\lambda=0$ and then use the implicit function theorem to
show that there exists a solution for small $\lambda$.

\subsect{lz}{$\lambda = 0$}

Fix $R >0$, assume $p>3$, $-1<\delta<-p/3$, $\mu>0$, and let
$\Lambda >0$ be as in proposition \ref{smoothD}. Then
the expansion from proposition \ref{smoothC} shows
that 
\eqn{rmeqnsA}{
\Xi(0,\U,\Psi,Z) = 0
}
if and only if
\lalign{rmeqnsB}{
& \Delta \U^{0\alpha}  = 0  \, , \label{rmeqnsB.1} \\
& \Delta \U^{ij} = \delta^{ik}\delta^{jl}\partial_{k}\U^{00}\partial
_{l}\U^{00}- 
\Half \delta^{ij}\delta^{kl}
\partial_{k}\U^{00}\partial_{l}\U^{00} + 4\pi G \To{}^{ij}(\Psi,Z) \, , \label{rmeqnsB.2} \\
& \delta^{ij}\Do{}^{A}_{i}\Do{}^{A}_{j}\Phi = 0 \, , \label{rmeqnsB.3} \\
& \delta^{kl}\Do{}^{A}_{k}F^{A}_{li} -\Bigl(\chi_{1}[\Phi,\Do{}^{A}_{
i}\Phi]
+(1-\chi_{1})|\Phi|^2\Bigl(\frac{\chi_{1}}{r^{2}}
\epsilon_{i}{}^{j}{}_{k}x^{k}\tau_{j}+ Z_{i}\Bigr)\Bigr) \, , \label{rmeqnsB.4}
}
where $A=Y+Z$ and $\Phi = \Omega + \Psi$. Equations \eqref{rmeqnsB.1}-
\eqref{rmeqnsB.4} can be regarded as the Newtonian YMH equations with
$\U^{00}$ playing the role of the Newtonian potential.

The BPS monopole solution to the Yang-Mills-Higgs equation is
\leqn{BPSa}{
A^{b} = \frac{w-1}{r^2}\epsilon_{i}{}^{j}{}_{k}x^{k}\tau_{j} \quad
\text{and} \quad \Phi^{b} = \frac{\phi(r)}{r}x^{j}\tau_{j}
}
where
\leqn{BPSb}{
w(r) = \frac{r}{\sinh(r)} \quad \text{and} \quad \phi(r) = \coth(r)-\frac{1}{r} \, .
}
From this we define
\leqn{BPSc}{
Z^{b} := A^{b}-Y= \frac{w-\chi_{1}}{r^2}\epsilon_{i}{}^{j}{}_{k}x^{k}\tau_{j} i\, , \quad 
\Psi^{b} := \Phi^{b}-\Omega = \frac{(\phi(r)-1)+\chi_{1}}{r}x^{j}\tau_{j}\, ,
}
and also observe that 
\eqn{BPSd}{
Z^{b} \in \Ac^{k,p}_{\mu} \quad \text{and} \quad
\Psi^{b} \in \Hc^{k,p}_{\delta} \quad \text{for $1\leq p \leq \infty$, $k\in \Nbb_{0}$ and $\delta, \mu > -1$.}
}
It can be checked that $(\Psi^{b},Z^{b})$ solve equations \eqref{rmeqnsB.3} and \eqref{rmeqnsB.4}.
Then using lemma \eqref{slaplaceB},  
\leqn{Ulz}{
\U_{b}^{0\alpha} := 0\, , \quad \U_{b}^{ij} := \Delta^{-1}\To{}^{ij}(\Psi^{b},Z^{b})
}
solve the remaining equations \eqref{rmeqnsB.1} and \eqref{rmeqnsB.2}\, .

\subsect{lgz}{$\lambda > 0$}

To use the implicit function theorem, we first need to establish that the derivative
of the map 
\eqn{Ximap0}{
\Xi_{0}(\U,\Psi,Z) := \Xi(0,\U,\Psi,Z) 
}
evaluated at $(\U_{b},\Psi^{b},Z^{b})$ is an isomorphism.

\begin{prop} \label{solA}\mnote{[solA]}
Suppose $p > 6$, $-1<\delta <-3/p$ and $-1<\mu < 0$. Then the linear map
\eqn{solA.1}{
D\Xi_{0}(\U_{b},\Psi^{b},Z^{b}) : \U^{2,p}_{\delta}\times
\Hc^{2,p}_{\delta}\times
\Ac^{2,p}_{\mu} \longrightarrow  \U^{0,p}_{\delta-2}\times
\Hc^{0,p}_{\delta-2}\times
\Ac^{0,p}_{\mu} 
}
is an isomorphism.
\end{prop}
\begin{proof}
For $\vPsi \in \sHo$, a short calculation shows that
\eqn{ident1}{
\delta^{ij}[A^{b}_{i},[A^{b}_{j},\vPsi]] = -2\frac{(w-1)^2}{r^2}\vPsi \quad 
\text{and}\quad \delta^{ij}[A^{b}_{i},\partial_{j}\vPsi] = -2\frac{(w-1)}{r^2}\vPsi \, .
}
This and \eqref{divA2} shows that
\eqn{ident2}{
\delta^{ij}\Do{}^{A^{b}}_{i}\Do{}^{A^{b}}_{j}\vPsi = \Delta \vPsi - 2\frac{w^{2}(r)-1}{r^2}\vPsi
\quad \text{for all $\vPsi \in \sHo$.}
}
But since $\sHo\times \sAo$ is dense in $\Hc^{2,p}_{\delta}\times \Ac^{2,p}_{\mu}$,
the continuity of the maps  
$\delta^{ij}\Do{}^{A^{b}}_{j}\Do{}^{A^{b}}_{i}:\Hc^{2,p}_{0}\rightarrow \Hc^{0,p}_{\delta-2}$
(see \eqref{DY}) and $ \Delta - 2 r^{-2}(w^{2}(r)-1) : \Hc^{2,p}_{0}\rightarrow \Hc^{0,p}_{\delta-2}$
(see proposition \eqref{slaplaceA}) implies that
\leqn{ident3}{
\delta^{ij}\Do{}^{A^{b}}_{i}\Do{}^{A^{b}}_{j}\vPsi = \Delta \vPsi - 2\frac{w^{2}(r)-1}{r^2}\vPsi
\quad \text{for all $\vPsi\in \Hc^{2,p}_{\delta}$.}
}
Using this and \eqref{divA2}, the derivative of 
$\Xi_{0}$ at $(\U^{b},\Psi^{b},Z^{b})$ can be written as 
\lalign{deriv1a}{
D\Xi(0,\Psi^{b},Z^{b})\cdot\begin{pmatrix}\vU 
\\ \vPsi \\ \vZ \end{pmatrix} &= \begin{pmatrix}
\Delta & 0 & 0 \\
0 & \Delta - 2\frac{w^2-1}{r^2} & 0\\
0 & 0 & \Delta - 1 
\end{pmatrix} \begin{pmatrix}\vU \\ \vPsi \\ \vZ \end{pmatrix}
\notag \\
&\quad +\begin{pmatrix} 0 & J_{1} & J_{2} \\ 
0 & 0 & K_{12} \\ 0 & K_{21} & K_{22}  \end{pmatrix} 
\begin{pmatrix}\vU \\ \vPsi \\ \vZ \end{pmatrix} \, ,
\label{deriv1}} 
where
\lalign{deriv2}{
J_{1}(\vPsi)^{ij} & = 2\delta^{ik}\delta^{jl}
\ip{\Do{}^{A^{b}}_{k}\Phi^{b}}{\Do{}^{A^{b}}_{l}\vPsi} - \delta^{ij}\delta^{kl}
\ip{\Do{}^{A^{b}}_{k}\Phi^{b}}{\Do{}^{A^{b}}_{l}\vPsi} \, , \label{deriv2.1} \\
J_{2}(\vZ)^{ij} & = \ip{\Do{}^{A^{b}}_{k}\Phi^{b}}{[\vZ_{l},\Phi^{b}]} - \delta^{ij}\delta^{kl}
\ip{\Do{}^{A^{b}}_{k}\Phi^{b}}{[\vZ_{l},\Phi^{b}]} + \notag \\
& \qquad 2\ip{F^{A^{b}}_{km}}{\vF_{ln}}-\Half \delta^{kl}\delta^{mn}\delta^{ij}
\ip{F^{A^{b}}_{km}}{\vF_{ln}} \, , \label{deriv2.2}\\
K_{12}(\vZ) & = \delta^{ij}\bigl( \partial_{i}[\vZ_{j},\Phi^{b}] + [\vZ_{i},\Do{}^{A^{b}}_{j}\Phi^{b}]
+[A^{b}_{i},[\vZ_{j},\Phi^{b}]] \bigr) \, , \label{deriv2.3} \\
K_{21}(\vPsi)_{i} &= -\chi_{1}\bigl([\vPsi,\Do{}^{A^{b}}_{k}\Phi^{b}]+
[\Phi^{b},\Do{}^{A^{b}}_{l}\vPsi]\bigr)\notag \\
&-(1-\chi)2\ip{\Phi^{b}}{\vPsi}
\Bigl(\frac{\chi_{1}}{r^2}\epsilon_{i}{}^{j}{}_{k}x^{k}\tau_{j}+Z^{b}_{i}\Bigr) \, , \label{deriv2.4} \\
K_{22}(\vZ)_{i} & = \delta^{kl}\Bigl(\partial_{k}\bigl([\vZ_{l},A^{b}_{i}]+[A^{b}_{l},\vZ_{i}]\bigr)
+[\vZ_{k},F^{A^{b}}_{li}]+[A^{b}_{k},\vF_{li}]\Bigr) \notag \\
& \qquad -\chi[\Phi^{b},[\vZ_{i},\Phi^{b}]]-\bigl( (|\Phi^{b}|^2-1)-\chi_{1}|\Phi^{b}|^2\bigr)\vZ_{i}
\, , \label{deriv2.5}
} 
and
\eqn{varF}{
\vF_{li} = \partial_{l}\vZ_{i}-\partial_{i}\vZ_{l} +[\vZ_{l},A^{b}_{i}]+[A^{b}_{l},\vZ_{i}] \, .
}

Since $\Delta : \Uc^{2,p}_{\delta} \rightarrow \Uc^{0,p}_{2}$ is an isomorphism (see proposition
\ref{slaplaceB}), it follows from the structure of the \eqref{deriv1} that 
$D\Xi_{0}(\U_{b},\Psi^{b},Z^{b})$ will be an isomorphism provided that 
\eqn{solA.3}{
S:= \begin{pmatrix} 
\Delta - 2\frac{w^2-1}{r^2} & 0\\
0 & \Delta - 1 \end{pmatrix}
+\begin{pmatrix} 0 & K_{12} \\ K_{21} & K_{22}  \end{pmatrix} 
} 
is an isomorphism. Let  
\eqn{Kmap}{
K = \begin{pmatrix} 0 & K_{12} \\ K_{21} & K_{22} \end{pmatrix}
}
Then  the weighted Rellich-Kondrachov theorems (see lemma \ref{esoblem2} and lemma 2.1 
of \cite{CHCH81}),  lemma \ref{esoblem3}, theorem 1.2 (iv) of \cite{Bart86}, and the inclusion
\eqref{mudel} shows that map $K: \Hc^{2,p}_{\delta}\times \Ac^{2,p}_{\mu}\rightarrow \Hc^{0,p}_{\delta-2}
\times \Ac^{0,p}_{\mu}$ is compact. As the Index of a operator is preserved under compact
perturbations, we get 
\leqn{solA.5}{ 
\text{Index}\bigl(S) = 0 
}
by  propositions \ref{slaplaceA} and \ref{sYukawa}. Thus if we can establish that $S$ is
injective then the proof will be complete.

\begin{lem} \label{Sinj} \mnote{[Sinj]}
\eqn{solA.6}{
\text{\emph{Ker}}(S) = {0}
}
\end{lem}
\begin{proof}
We first consider the YMH Lagrangian   
\leqn{Lag}{
L(\Psi,Z) = \int_{\Rbb^{3}}\Half\delta^{ik}\delta^{ij}\ip{F^{A}_{ik}}{F^{A}_{jl}}
+ \delta^{ij}\ip{\Do{}^{A}_{i}\Phi}{\Do{}^{A}_{j}\Phi} d^{3}x\, .
}
where $A=Y+Z$ and $\Phi = \Omega+\Psi$ as above. Since $p>6$ and $-1 < \delta <-3/p$, 
we get from \eqref{Fmap}, \eqref{DY}, theorem 2.1 (i) of \cite{Bart86} 
and the multiplication lemma \ref{multiply}
that the map 
\eqn{solA.7}{
\Hc^{2,p}_{\delta}\times \Ac^{2,p}_{\mu} \rightarrow
\W^{0,p}_{2\delta-2}(\Rbb^{3}) \subset L^{1}(\Rbb^{3})\; :\; (\Psi,Z)
\rightarrow \Half\delta^{ik}\delta^{ij}\ip{F^{A}_{ik}}{F^{A}_{jl}}
+ \delta^{ij}\ip{\Do{}^{A}_{i}\Phi}{\Do{}^{A}_{j}\Phi}
}
is analytic. Consequently the Lagrangian \eqref{Lag} defines analytic map from
$\Hc^{2,p}_{\delta}\times \Ac^{2,p}_{\mu}$ to $\Rbb$. Differentiating \eqref{Lag} yields
\lalign{DLag}{
DL(\Psi,Z)&\cdot(\vPsi,\vZ)   =
\int_{\Rbb^{3}} \delta^{ik}\delta^{ij}\ip{F^{A}_{ik}}{\partial_{j}\vZ_{k}-\partial_{k}\vZ_{j}
+[\vZ_{j},A_{k}]+[A_{j},\vZ_{k}]} d^{3}x \notag \\
& \quad + \int_{\Rbb^{3}} 2\delta^{ij}\ip{\Do{}^{A}_{i}\Phi}{\Do{}^{A}_{j}\vPsi+[\vZ_{j},\Phi]} 
d^{3}x \notag \\
& = -\int_{\Rbb^{3}}2\delta^{ik}\delta^{jl}\ip{\Do{}^{A}_{i}F_{kj}-[\Phi,\Do{}^{A}_{j}\Phi]}{\vZ_{l}}
+ 2\delta^{ij}\ip{\Do{}^{A}_{i}\Do{}^{A}_{j}\Phi}{\vPsi} d^{3}x  \notag
}
where in deriving the last inequality we used integration by parts.
A similar calculation
shows that the second derivative evaluated on the diagonal is
\lalign{D2LagA}{
D^{2}L(\Psi,Z)\cdot\bigl((\vPsi,\vZ),(\vPsi,\vZ)\bigr)
=&-2\int_{\Rbb^{3}}\delta^{ij}\ip{L_{2}(\Psi,Z)\cdot(\vPsi,\vZ)_{i}}{\vZ_{j}}
\notag \\
&+ \ip{L_{1}(\Psi,Z)\cdot (\vPsi,\vZ)}{\vPsi} d^{3}x
\label{D2Lag}}
where 
\lgath{LYMH}{
L_{1}(\Psi,Z)\cdot(\vPsi,\vZ) = \delta^{ij}\Bigl(\Do{}^{A}_{i}\bigl(\Do{}^{A}_{j}\vPsi 
+[\vZ_{i},\Phi] \bigr) + [\vZ_{i},\Do{}^{A}_{j}\Phi] \Bigr)\, , \label{LYMH.1}\\
L_{2}(\Psi,Z)\cdot(\vPsi,\vZ)_{j} = \delta^{ik}\Bigl(\Do{}^{A}_{i}\bigl(
\partial_{k}\vZ_{j}-\partial_{j}\vZ_{j} + [\vZ_{k},A_{j}],[A_{k},\vZ_{j}]
\bigr) + [\vZ_{i},F^{A}_{kj}]\Bigl) \notag \\
 - [\Psi,\Do{}^{A}_{j}\Phi]-
[\Phi,\Do{}^{A}_{j}\Psi+[\vZ_{j},\Phi]] \, . \label{LYMH.2}
}

Let
\eqn{starF}{
*F^{A}_{k} = \Half\epsilon^{ij}{}_{k}F^{A}_{ij}
}
be the Hodge dual of $F^{A}$. Then the Bianchi identities for $F^{A}$ imply that
\lgath{mono1}{
\delta^{ij}\ip{*F^{A}_{i}+\Do{}^{A}_{i}\Phi}{*F^{A}_{j}+\Do{}^{A}_{j}\Phi}dx^{1}\wedge dx^{2}\wedge dx^{3}
-d\bigl(\ip{\Phi}{F^{A}_{ij}}dx^{i}\wedge dx^{j}\bigr) = \notag \\
\bigl(\Half\delta^{ik}\delta^{ij}\ip{F^{A}_{ik}}{F^{A}_{jl}}
+ \delta^{ij}\ip{\Do{}^{A}_{i}\Phi}{\Do{}^{A}_{j}\Phi}\bigr)dx^{1}\wedge dx^{2}\wedge dx^{3} \, . \notag 
} 
Therefore the Lagrangian \eqref{Lag} can be written as
\eqn{aLag1}{
L(\Psi,Z) = \int_{\Rbb^{3}}\delta^{ij}\ip{*F^{A}_{i}+\Do{}^{A}_{i}\Phi}{*F^{A}_{j}+\Do{}^{A}_{j}\Phi} d^{3}x
-\int_{\Rbb^{3}} d\bigl(\ip{\Phi}{F^{A}_{ij}}dx^{i}\wedge dx^{j}\bigr) \, .
}
But for $(\Psi,Z)\in \sHo \times \sAo$ we have
we have that
\lgath{surface1}{
\int_{\Rbb^{3}} d\bigl(\ip{\Phi}{F^{A}_{ij}}dx^{i}\wedge dx^{j}\bigr)  = 
\int_{\Rbb^{3}}d\bigl(\ip{\Omega}{F^{Y}_{ij}}dx^{i}\wedge dx^{j}\bigr)\notag \\
+ \int_{\Rbb^{3}}d\bigl(\ip{\Omega}{ F^{Z}_{ij}+[Z_{i},Y_{j}]+[Y_{i},Z_{j}]}
+\ip{\Psi}{F^{Y}_{ij}+ F^{Z}_{ij}+[Z_{i},Y_{j}]+[Y_{i},Z_{j}]}dx^{i}\wedge dx^{j}\bigr) \notag
\\
= \lim_{R\rightarrow\infty} \int_{\partial B_{R}(0)}\ip{\Omega}{F^{Y}_{ij}}dx^{i}\wedge dx^{j}
= 4\pi
}
where we have used Stokes' theorem to convert to a surface integral.
Using the weighted Sobolev inequalities (see theorem \ref{esobthm1} and theorem 1.2 of \cite{Bart86}),
it follows from the density of $\sHo \times \sAo$ that
\eqn{surface2}{
\int_{\Rbb^{3}} d\bigl(\ip{\Phi}{F^{A}_{ij}}dx^{i}\wedge dx^{j}\bigr) =  4\pi \; 
\text{ for all $(\Psi,Z)\in  \Hc^{2,p}_{\delta}\times  \Ac^{2,p}_{\mu}$.}
}
Thus we have the alternate form for the Lagrangian
\eqn{aLag2}{
L(\Psi,Z) = \int_{\Rbb^{3}}\delta^{ij}\ip{*F^{A}_{i}+\Do{}^{A}_{i}\Phi}{*F^{A}_{j}+\Do{}^{A}_{j}\Phi} d^{3}x
-4\pi .
}
This way of expressing the Yang-Mills-Higgs Lagrangian is well
known and leads to  Bogomol'nyi first order equations.
Differentiating the above Lagrangian  twice and using integration by parts yields
\lgath{aD2Lag}{
D^{2}L(\Psi,Z)\cdot\bigl((\vPsi,\vZ),(\vPsi,\vZ)\bigr)
 = \int_{\Rbb^{3}} 2\delta^{ij}\ip{*F^{A}_{i}+\Do{}^{A}_{i}\Phi}{M_{1}(\vPsi,\vZ)_{j}}
+\notag  \\
 2\delta^{ij}\ip{M_{2}(\Psi,Z)\cdot(\vPsi,\vZ)_{i}}{M_{2}(\Psi,Z)\cdot(\vPsi,\vZ)_{j})}
d^{3}x \label{aD2Lag.1}
}
where
\lalign{aLYMH}{
M_{1}(\vPsi,\vZ)_{k} & = \epsilon^{ij}{}_{k}[\vZ_{i},\vZ_{j}] + 2[\vZ_{k},\vPsi] \label {aLYMH.1} \\
M_{2}(\Psi,Z)\cdot(\vPsi,\vZ)_{k} & = \Half\epsilon^{ij}{}_{k}\bigl( \partial_{i}\vZ_{j}
-\partial_{j}\vZ_{i} + [\vZ_{i},A_{j}]+[A_{i},\vZ_{j}]\bigr) \notag \\
& +
\bigl(\Do{}^{A}_{k}\vPsi+[\vZ_{k},\Phi]\bigr) \, . \label{aLYMH.2}
}

Now suppose that $(\vPsi,\vZ) \in  \Hc^{2,p}_{\delta}\times  \Ac^{2,p}_{\mu}$ satisfies
$S(\vPsi,\vZ) = 0$. Since $S$ is an elliptic operator with smooth coefficients, elliptic regularity
implies that  $\vPsi$ and $\vZ$ are $C^{\infty}$. Then using \eqref{yh5} and \eqref{ident3} 
$\vPsi$ and $\vZ$ satisfy
\eqn{solA.11}{
L_{1}(\Psi^{b},Z^{b}) \cdot (\vPsi,\vZ) = 0 \quad \text{and} \quad
L_{2}(\Psi^{b},Z^{b}) \cdot (\vPsi,\vZ) = 0 \, .
}
Also, we  note that $\Phi^{b}$ and $A^{b}$ satisfy the 
Bogomol'nyi  equations
\eqn{solA.12}{
*F^{A^{b}}_{j} + \Do{}^{A^{b}}_{j}\Phi^{b} = 0 \quad \Longleftrightarrow \quad
w'+w\phi=0, \quad r^2\phi'+w^2-1 = 0 \, .
}
So we get by \eqref{D2Lag} and \eqref{aD2Lag.1} that
\leqn{solA.13}{
\Half\epsilon^{ij}{}_{k}\bigl( \partial_{i}\vZ_{j}
-\partial_{j}\vZ_{i} + [\vZ_{i},A^{b}_{j}]+[A^{b}_{i},\vZ_{j}]\bigr) +
\bigl(\Do{}^{A^{b}}_{k}\vPsi+[\vZ_{k},\Phi^{b}]\bigr) = 0 \, .
}
Letting $\vPsi = \psi(r)r^{-1}x^{k}\tau_{k}$ and $\vZ_{i} = z(r)r^{-2}\epsilon_{i}{}^{j}_{k}x^{k}\tau_{j}$,
we can write \eqref{solA.13} as
\leqn{solA.14}{
z'+\phi z+w\psi = 0 \quad \text{and} \quad r^2\psi'+2wz=0 \, ,
}
where $w(r)$ and $\phi(r)$ are given by \eqref{BPSb}. Differentiating $1/w$ times
the second equation  and then using the two equations to eliminate  $z$ and $z'$  yields
\leqn{solA.15}{
(r^2\psi')'+2\phi r^{2}\psi' -2w^2\psi = 0 \, .
}
Since $\vPsi\in C^{\infty}\cap \Ac^{2,p}_{\mu}$  ($-1<\delta < 0$) 
we get that $\psi(r)=\Ord(r)$ as $r\rightarrow 0$ and that
$\psi(r)=\ord(r^{\delta})$ as $r\rightarrow \infty$ by theorem 1.2 of \cite{Bart86}.
Since $w>0$ on $[0,\infty)$ the only solution satisfying the
 differential equation \eqref{solA.15} and the asymptotic conditions
is the trivial solution $\psi=0$. But $\psi=0$ implies that $z=0$ and thus
$\vPsi=0$ and $\vZ=0$.  This establishes that $\text{Ker}(S)$ is
trivial.
\end{proof}
\end{proof}

We can now solve the reduced/modified EYMH equations.
\begin{thm} \label{solB}\mnote{[solB]}
Suppose $p > 6$, $-1<\delta <-3/p$ and $-1< \mu < 0$. Then there exists
a $\Lambda > 0$ and an analytic map
\eqn{solB.1}{
(-\Lambda,\Lambda)\longrightarrow  \U^{2,p}_{\delta}\times
\Hc^{2,p}_{\delta}\times \Ac^{2,p}_{\mu} \: :\: \lambda \longmapsto (\U(\lambda),\Psi(\lambda),Z(\lambda))
}
such that $(\U(0),\Psi(0),Z(0))= (\U_{b},\Psi^{b},Z^{b})$
and $\Xi(\lambda,\U(\lambda),\Psi(\lambda),Z(\lambda)) = 0$ for all $\lambda \in (-\Lambda,\Lambda)$.
\end{thm}
\begin{proof}
Propositions \ref{smoothD} and \ref{solA} and the results of section \ref{lz}, allow
us to apply the analytic version of the implicit function theorem (see \cite{Deim} theorem 15.3)
to reach the desired conclusion.
\end{proof}

%% file: exist.tex
\sect{exist}{Existence}

We have so far only found a solution to the reduced/modified EYMH equations
\eqref{reduced3}, \eqref{yh6b.1}, and \eqref{yh7.1}. However, we will now show
that the solution obtained in theorem \ref{solB} is also a solution to
the EYMH equations \eqref{YMHa1}-\eqref{YMHa2}.

\begin{prop} \label{existA} \mnote{[existA]}
Suppose $p > 6$, $-1<\delta <-3/p$, and $-1<\mu < 0$. Let
\eqn{existA.1}{
(-\Lambda,\Lambda)\longrightarrow  \U^{2,p}_{\delta}\times
\Hc^{2,p}_{\delta}\times \Ac^{2,p}_{\mu} \: :\: \lambda \longmapsto (\U(\lambda)
,\Psi(\lambda),Z(\lambda))
}
be the map from theorem \ref{solB}. Then there exists a $\Lambda^{*}\in (0,\Lambda]$
such that for every $\lambda \in (-\Lambda^{*},\Lambda^{*})$,
$\bigl(\U(\lambda) ,\Phi(\lambda)=\Omega+\Psi(\lambda),A(\lambda)=Y+Z(\lambda)\bigr)$ solves the YMH equations \eqref{YMHa1}-\eqref{YMHa2}
and $(\Psi(\lambda),Z(\lambda)) \in \Uc^{2,p}_{\delta}\cap C^{1}\times\Hc^{2,p}_{\delta}\cap 
C^{2}\times\Ac^{2,p}_{\mu}\cap C^{2}$. 
\end{prop}
\begin{proof}
Fix $R > 0$. Then for each $\lambda \in (-\Lambda,\Lambda)$, 
$\U(\lambda)\in \W^{2,p}(B_{R}(0),\Sbb^{3})$, $\Psi \in \W^{2,p}(B_{R}(0),\sU{2})$,
and $Z(\lambda) \in \W^{2,p}(B_{R}(0),\sU{2}^{3})$. To reduce notation we will
often write $\U$, $\Psi$, and $Z$ instead of
$\U(\lambda)$, $\Psi(\lambda)$, and $Z(\lambda)$. Since $Y$ and $\Omega$ are
$C^{\infty}$ it follows from \eqref{ximap.2}-\eqref{ximap.3} and the Sobolev
inequalities that
\eqn{existA.2}{
\gb^{ij}\partial^{2}_{ij}\Psi = f \quad \text{and} \quad Q^{ijk}{}_{l}\partial^{2}_{ij}Z_{k} = h_{l}
} 
where $f,h_{l} \in \W^{1,p}(B_{R}(0),\sU{2}) \subset C^{0,1-3/p}(B_{R}(0),\sU{2})$ 
and
\gath{existA.3}{
\gb^{ij} = \delta^{ij} + 4\lambda^{2}\U^{ij} \, , \\
Q^{ik} = \left(Q^{ikl}{}_{j}\right) := \big( (\delta^{ik}+4\lambda^{2}\U^{ik})
\delta^{l}_{j} - 4\lambda^{2}\U^{lk}\delta^{i}_{j}\big) \, 
}
By the weighted Sobolev inequality, \cite{Bart86} theorem 1.2 (v), the
embedding $\W^{1,p}_{\delta}(\Rbb^{3},\Sbb^{3}) \rightarrow \C^{0,1-3/p}_{\delta}(\Rbb^{3},\Sbb^{3})$ is
continuous and hence the map $(-\Lambda,\Lambda) \rightarrow  
\C^{0,1-3/p}_{\delta}(\Rbb^{3},\Sbb^{3}) \: : \: \lambda \mapsto U(\lambda)$ is continuous. Therefore, 
there exists a $\Lambda^{*} \in (0,\Lambda)$ such that
the operators
$\gb^{ij}\partial^{2}_{ij}$ and $Q^{ij}\partial^{2}_{ij}$
are uniformly elliptic with with coefficients in   $\C^{0,1-3/p}_{\delta}(\Rbb^{3})$ 
for all $\lambda \in [-\Lambda^{*},\Lambda^{*}]$. By
elliptic regularity, $\Psi,Z_{k}\in C^{2}(B_{R}(0),\sU{2})$. As $\Lambda^{*}$ is independent
of $R$, we get that  $\Psi(\lambda),Z_{k}(\lambda)\in C^{2}(\Rbb^{3},\sU{2})$ for
all $\lambda \in (-\Lambda^{*},\Lambda^{*})$.

For $\lambda > 0$ we can,
using \eqref{metfromdens1}, recover the metric $g_{\alpha\beta}$ from $\U_{\alpha\beta}$.
Since $\U\in \W^{2,p}_{\delta}(\Rbb^{3},\Sbb)$, we have by theorem 1.2 (v) of \cite{Bart86}
that $\U^{\alpha\beta}\in  \C^{0,1-3/p}_{\delta}(\Rbb^{3})$ and
$\partial_{k}\U^{\alpha\beta} \in \C^{0,1-3/p}_{\delta-1}(\Rbb^{3})$. Therefore, in
spherical coordinates the metric becomes
\eqn{existA.4}{
g_{\alpha\beta} dx^{\alpha}dx^{\beta} =
-S(r)N(r)dt^2 + \frac{1}{N(r)}dr^2 + R(r)^2(d\theta^2+\sin^{2}\theta d\phi^{2}) \, ,
}
where $N$, $S$, and $R$ are in $C^{1}((0,\infty))$. But then a straightforward
calculation shows that for all $r\in (0,\infty)$
\alin{existA.5}{
(1-&\chi_{3})g^{\alpha\beta}\D{}^{Y}_{\alpha}F^{Y}_{\beta \nu}dx^{\nu}
= \\& (1-\chi_{3})\Bigl(\frac{1}{S}(NS\chi_{1}')'-\frac{(\chi_{1}^2-1)\chi_{1}}{R^2}\Bigr)
\bigl(-\sin\phi\,\tau_{1}+ 
\cos\phi\, \tau_{2}\bigr)d\theta \\
& +(\chi_{3}-1)\Bigl(\frac{1}{S}(NS\chi_{1}')'-\frac{(\chi_{1}^2-1)\chi_{1}}{R^2}\Bigr)
\bigl(\tau_{3} 
-\cot\theta(\sin\phi\,\tau_{2}+\cos\phi\,\tau_{1})\bigr)d\phi = 0 
\, .
}
where $(\cdot)'=\frac{d\,}{dr}(\cdot)$. Using this result and the identity \eqref{yh5}
which is valid for $C^{1}$ static spherically symmetric fields, it is clear
that $(\U,\Psi,Z)$ satisfy the YMH equations.
\end{proof}

To complete the existence proof, we now use the following result of Heilig.
\begin{prop}{\emph{[proposition 6.1,\cite{Heil95}]}} \label{existB} \mnote{[existB]}
Suppose $-1<\delta < 0$, $p>3$, and $\Lambda > 0$. Furthermore, suppose
\gath{existB1}{
T : [0,\Lambda] \rightarrow \emph{\W}^{0,p}_{\delta-2}(\Rbb^{3},\Sbb^{3}) \cap
\emph{\text{C}}^{1}(\Rbb^{3},\Sbb^{3}) \; : \; \lambda \mapsto
(T^{\alpha\beta}_{\lambda} )
\intertext{and}
\U : [0,\Lambda] \rightarrow  \emph{\W}^{2,p}_{\delta}(\Rbb^{3},\Sbb^{3}) \;
: \; \lambda \mapsto (\U^{\alpha\beta}_{\lambda})
}
are two continuous maps such that for every $\lambda \in [0,\Lambda]\;$:
$(\lambda,\U^{\alpha\beta}_{\lambda},T^{\alpha\beta}_{\lambda})$
is a solution to the reduced field equations \ref{reduced2},
$\nabla_{\beta} T^{\alpha\beta}_{\lambda} = 0$, and
$\partial_{\gamma} T^{\alpha\beta}_{\lambda}
\in B_{\emph{\W}^{0,p}_{\delta-2}(\Rbb^{3})}(0,R)$ for some $R > 0$ independent
of $\lambda$ and $\alpha,\beta,\gamma$. Then there exists a constant $\hat{\Lambda} \in
(0,\Lambda]$ such that
$\partial_{\alpha} \U^{\alpha\beta}_{\lambda} = 0$ for all
$\lambda \in [0,\hat{\Lambda}]$.
\end{prop}

\begin{thm} \label{existC} \mnote{[existC]}
Suppose $p > 6$, $-1<\delta <-3/p$, and $-1<\mu < 0$. Let
\eqn{existC.1}{
(-\Lambda,\Lambda)\longrightarrow  \U^{2,p}_{\delta}\times
\Hc^{2,p}_{\delta}\times \Ac^{2,p}_{\mu} \: :\: \lambda \longmapsto (\U(\lambda)
,\Psi(\lambda),Z(\lambda))
}
be the map from theorem \ref{solB}. Then there exists a $\Lambda^{*}\in (0,\Lambda]$
such that for every $\lambda \in (-\Lambda^{*},\Lambda^{*})$,
$\bigl(\U(\lambda) ,\Phi(\lambda)=\Omega+\Psi(\lambda),A(\lambda)=Y+Z(\lambda)\bigr)$ solves the EYMH equations \eqref{reduced2}-\eqref{reduced3}
and \eqref{YMHa1}-\eqref{YMHa2}. Moreover,
$(\U(\lambda),\Psi(\lambda),Z(\lambda)) \in \Uc^{2,p}_{\delta}\cap C^{2}\times\Hc^{2,p}_{\delta}\cap 
C^{2}\times\Ac^{2,p}_{\mu}\cap C^{2}$ for all $\lambda \in (-\Lambda^{*},\Lambda^{*})$. 
\end{thm}
\begin{proof}
From proposition \ref{existA} we know that there exist a $\Lambda^{*}\in (0,\Lambda]$ such that
$\bigl(\U(\lambda) ,\Phi(\lambda)=\Omega+\Psi(\lambda),A(\lambda)=Y+Z(\lambda)\bigr)$
solves the YMH equations  \eqref{YMHa1}-\eqref{YMHa2}. and  $\U(\lambda)\in C^{1}(\Rbb^{3},\Sbb^{3})$
,$\Psi(\lambda),A_{k}(\lambda)\in C^{2}(\Rbb^{3},\sU{2})$ for all $\lambda \in (-\Lambda^{*},\Lambda)$.
It can then be checked that the YMH equations imply that  $\nabla_{\alpha} T^{\alpha\beta} = 0$ is
automatically satisfied. Therefore, the harmonic equation 
\leqn{existC.3}{
\partial_{\alpha}\U^{\alpha\beta}=0
}
is satisfied for all $\lambda \in (-\Lambda^{*},\Lambda)$ by propositions \ref{smoothB} and 
\ref{existB}. So we have shown that $\bigl(\U(\lambda) ,\Phi(\lambda)=\Omega+\Psi(\lambda),
A(\lambda)=Y+Z(\lambda)\bigr)$ satisfies the EYMH equations
\eqref{reduced2}-\eqref{reduced3}
and \eqref{YMHa1}-\eqref{YMHa2} for all  $\lambda \in (-\Lambda^{*},\Lambda)$. To complete the
the proof we use \eqref{existC.3} to write the reduced equations \eqref{reduced3} as
\eqn{existC.4}{
\gb^{ij}\partial^{2}_{x^{i}x^{j}}\U^{\alpha\beta} = H^{\alpha\beta}
}
where $H^{\alpha\beta} = -A^{\alpha\beta} - B^{\alpha\beta} - C^{\alpha\beta} +
4\pi G |\df| T^{\alpha\beta}$. As in proposition \ref{existA}, it can be shown that
there exist a $\hat{\Lambda}>0$ such that for all
$\lambda \in (0,\hat{\Lambda})$ and
$R > 0$ that $H^{\alpha\beta} \in \text{C}^{0,1-3/p}(\text{B}_{R})$ and
the operator $\gb^{ij}\partial^{2}_{ij}$ is uniformly elliptic
with coefficients in $C^{0,1-3/p}_{\delta}(\Rbb^{3})$.
Therefore we conclude via elliptic regularity
that $\U^{\alpha\beta} \in \text{C}^{2}$.
\end{proof}
As the  Newtonian solutions \eqref{BPSc} and \eqref{Ulz} are $C^{\infty}$, we do not have to restrict
the  differentiability to $k=2$. All the same arguments go through for $k\geq 2$. Then using
the weighted Sobolev inequalities we get the following result: 

\begin{cor} \label{existD} \mnote{[existD]}
Suppose$-1<\delta < 0$, and $-1<\mu < 0$, $0<\alpha < 1$. Then for any integer $k \geq 2$ there
exist a constant $\Lambda > 0$ and an analytic map
\eqn{existD.1}{
(-\Lambda,\Lambda)\longrightarrow  
C^{k,\alpha}_{\delta}(\Rbb^{3},\Sbb) \times C^{k,\alpha}_{\delta}(\Rbb^{3},\sU{2})
\times  \Cc^{k,\alpha}_{\mu}(\Rbb^{3},\sU{2}^{3})\ \: :\: \lambda \longmapsto (\U(\lambda)
,\Psi(\lambda),Z(\lambda))
}
such that for every $\lambda \in (-\Lambda,\Lambda)$,
$\bigl(\U(\lambda) ,\Phi(\lambda)=\Omega+\Psi(\lambda),A(\lambda)=Y+Z(\lambda)\bigr)$ 
solves the EYMH equations  
\eqref{reduced2}-\eqref{reduced3} and $\bigl(\U(0)=\U_{b} ,\Phi(0)=\Phi^{b},A(0)=A^{b}\bigr)$.
\end{cor}